\documentclass[journal]{IEEEtran}

\ifCLASSINFOpdf
\else
\fi

\usepackage{hyperref}        
\usepackage{url}             
\usepackage{booktabs} 
\usepackage{amsfonts} 
\usepackage{nicefrac}        
\usepackage{microtype}       
\usepackage{bookmark}
\usepackage{array}
\usepackage{graphicx}
\usepackage{amsmath}
\usepackage{amsthm}
\usepackage{amsfonts} 
\usepackage{amssymb}
\usepackage{array}
\usepackage{algorithmic}
\usepackage{algorithm}
\usepackage{tabularx}
\usepackage{subfigure}
\usepackage{color}
\usepackage{letltxmacro}
\usepackage{cite}
\usepackage{multirow}
\usepackage{xcolor}
\usepackage{enumitem}
\usepackage{bm}

\usepackage[T1]{fontenc} 
\usepackage{amsmath}
\usepackage[cmintegrals]{newtxmath}
\usepackage{bm} 
\hypersetup{hidelinks}

\renewcommand{\baselinestretch}{1}

\usepackage{multirow}

\newtheorem{remark}{Remark}

\definecolor{airforceblue}{rgb}{0.36, 0.54, 0.66}
\definecolor{applegreen}{rgb}{0.55, 0.71, 0.0}
\definecolor{bittersweet}{rgb}{1.0, 0.44, 0.37}

\LetLtxMacro{\originaleqref}{\eqref}
\renewcommand{\eqref}{\originaleqref}

\newcommand{\rv}{\color{black}}

\title{\Huge Encoding Carbon Emission Flow in Energy Management: A Compact Constraint Learning Approach}

\author{
  Linwei~Sang,~\IEEEmembership{Graduate Student Member,~IEEE,}
  Yinliang~Xu$^*$,~\IEEEmembership{Senior Member,~IEEE,}
  Hongbin~Sun,~\IEEEmembership{Fellow,~IEEE}
  \thanks{Manuscript received Feb.~01, 2023; revised Apr.~07, 2023; accepted May.~04, 2023. This work was supported by the Shenzhen Science and Technology Program, Grant No. WDZC20220808143010001, and National Natural Science Foundation of China, Grant No. 52277107. (Corresponding Author: \emph{Yinliang Xu}).}
  \thanks{Linwei~Sang and Yinliang~Xu are with Tsinghua-Berkeley Shenzhen Institute, Tsinghua University, Shenzhen, China. (E-mail:\url{sanglinwei21@163.com}, \url{xu.yinliang@sz.tsinghua.edu.cn}).
  }
  \thanks{Hongbin Sun is with the Department of Electrical Engineering, State Key Laboratory of Power Systems, Tsinghua University, Beijing, China. (E-mail: \url{shb@tsinghua.edu.cn}).
  }
}

\begin{document}

\maketitle

\begin{abstract}
  Decarbonizing the energy supply is essential and urgent to mitigate the increasingly visible climate change. Its basis is identifying emission responsibility during power allocation by the carbon emission flow (CEF) model. However, the main challenge of CEF application is the intractable nonlinear relationship between carbon emission and power allocation. So this paper leverages the high approximation capability and the mixed-integer linear programming (MILP) representability of the deep neural networks to tackle the complex CEF model in carbon-electricity coordinated optimization. The compact constraint learning approach is proposed to learn the mapping from power injection to bus emission with sparse neural networks (SNNs). Then the trained SNNs are transformed equivalently as MILP constraints in the downstream optimization. In light of the ``high emission with high price'' principle, the blocked carbon price mechanism is designed to price emissions from the demand side. Based on the constraint learning and mechanism design, this paper proposes the carbon-aware energy management model in the tractable MILP form to unlock the carbon reduction potential from the demand side. The case study verifies the approximation accuracy and sparsity of SNN with fewer parameters for accelerating optimization solution and reduction effectiveness of demand-side capability for mitigating emission. 
\end{abstract}


\section{Introduction}

\IEEEPARstart{T}{O} tackle the increasingly visible climate change, the global energy systems should strive for carbon neutrality by around 2050 to limit the 1.5$^\circ$C global warming \cite{ Chen2021}. The rapidly growing energy demand makes the decarbonization of the energy supply essential and urgent \cite{Davis2018,Fu2023}. So effective engineering and innovative research are needed to pave the way toward a low-carbon energy society. Setting proper emission targets and reduction obligations is the key to mitigating carbon emissions effectively, and its basis lies in identifying the emission responsibility during the power allocation.

So carbon emission flow (CEF) model is proposed to clarify the carbon emission responsibility by calculating and analyzing the dynamic emission in the energy systems. It can trace the carbon emission accompanying the power allocation process from generation to demand, compared to the emission analysis on the generation side. Current researchers have utilized the CEF model to analyze the system emission for facilitating the system's low-carbon operation. Ref.~\cite{Kang2015} and \cite{Cheng2019} quantify the carbon emission in the power and multiple energy systems. A bi-level low-carbon energy management model is proposed in Ref.~\cite{Cheng2020} to reduce emissions efficiently based on the energy-carbon integrated prices from CEF. Ref.~\cite{Wang2020} investigates the carbon financing policies and their environmental benefits for consumers' response, where CEF aims to track carbon emissions. A decentralized multi-region energy planning model is proposed in Ref.~\cite{Cheng2019-tsg} to allocate the carbon emission cap based on the CEF for reducing emissions. {\rv The above research can calculate the CEF explicitly for the given optimal power flow results. But there rises new difficulty in the carbon-electricity coordinated optimization when both the carbon emission and power flow are coupled variables to be determined in a coordinated way.} The main challenge of CEF application is the intractable nonlinear relationship between carbon emission and power allocation, which cannot be transformed into tractable constraints {\rv due to numerous combinatorial bilinear terms}. So the above researches separate the calculation of CEF with the optimization of optimal energy dispatch for tractable calculation.

{\rv Fortunately, recent years have witnessed the burgeoning of machine learning applications to power system operation \cite{Zhang2022-Pro,Chen2022,Chatzos2022}, which provides powerful deep neural networks (DNNs) with the high representational capability. The duality theory is leveraged in Ref.~\cite{Chen2022} to propose a learning-based approach to predict the DCOPF results with theoretical characterizations of constraint satisfaction. A novel classification-plus-regression architecture is designed in Ref.~\cite{Chatzos2022} to predict the optimal solution of security-constrained economic dispatch. Different from the above learning-based power flow, we leverage the DNN to approximate the complex CEF model to achieve carbon-electricity coordinated optimization.} Ref.~\cite{Liang2017} demonstrates the predominant function approximation capability of DNN theoretically compared to shallow networks. The trained DNN with ReLU activation function can be exactly transformed into mixed-integer linear programming (MILP) \cite{Anderson2020}, embedded into the optimization model \cite{Maragno2021}, and solved efficiently by off-the-shelf optimizer \cite{Bertsimas2022}. Ref.~\cite{Zhang2022} utilizes the DNN to embed the intractable frequency constraints in preventive frequency-constrained unit commitment. The dense DNN in Ref.~\cite{Maragno2021, Bertsimas2022, Zhang2022, Kody2022} requires numerous parameters for approximation accuracy, so its equivalent MILP form will comprise many binary and continuous variables, which will impose a high calculation burden on the downstream {\rv efficient carbon-electricity coordinated optimization model}. {\rv However, recent ML research proposes the sparse neural networks with comparable learning capability and fewer parameters, which is widely applied for online prediction acceleration \cite{Frankle2019} and could be a possible solution to balance the learning accuracy and optimization efficiency, compared to full connected neural networks.}

This paper leverages the high approximation capability \cite{Liang2017} and the MILP representability \cite{Maragno2021} of DNNs to solve the carbon-aware energy management problem and proposes the compact constraint learning approach with the SNN to tackle the intractable CEF model. {\rv Different from the short-term emission prediction in Ref.~\cite{Huber2021} with the external information like load prediction, our proposed compact constraint learning focuses on learning the complex intractable CEF and encoding the trained SNN as constraints for the efficient carbon-electricity coordinated optimization.} It can learn the complex relationship from the power injection to the bus carbon emission through the offline CEF model in stochastic gradient descent (SGD) training and be encoded in the energy management model equivalently as the MILP constraints for online optimization. Compared with the dense DNN, SNN drops unnecessary connections iteratively for network sparsity in the sparse training process, leading to fewer parameters with higher calculation efficiency. Then, inspired by the time-of-use electricity price design \cite{Yang2019}, the blocked carbon price mechanism is designed to price the carbon emission from the demand side in light of the ``high emission with high price'' principle. Based on the compact constraint learning and the blocked carbon price mechanism, this paper proposes the data-driven carbon-aware energy management model to maximize the social welfare in the tractable MILP form to unlock the carbon reduction potential from the demand side. So the principal contribution of this paper can be summarized threefold:

1) To the authors' best knowledge, this paper, \emph{for the first time}, proposes the compact constraint learning approach to learn the complex relationship from the bus power injection to the bus carbon emission through SNN by leveraging its high approximation capability with fewer parameters and less complex constraints. {\rv Different from the emission prediction in Ref.~\cite{Huber2021} with the external information, our proposed approach focuses on learning CEF via SNN and encoding SNN-based CEF as constraints.}

2) Different from taking CEF as an indicator in Ref.~\cite{Cheng2019, Wang2020}, this paper incorporates the demand side carbon management into the energy management directly and designs the blocked carbon price mechanism, which can price the carbon emission from the demand side stepwise and unlock the carbon reduction potential from the load consumers, in light of the ``high emission with high price'' principle.

3) {\rv Inspired from the MILP representability of NN in Ref.~\cite{Anderson2020}, this paper proposes the SNN-based CEF in the MILP form based on the SNN learning of CEF and the equivalent MILP form of trained SNN. The trained SNN-based CEF is encoded in the energy management model with a carbon price mechanism to formulate the final data-driven carbon-aware energy management, which can incentivize the low-carbon operation of the power system and achieve carbon-electricity coordinated optimization.} The main novelty of this paper is the combination of advanced machine learning and mixed-integer optimization to achieve CEF-based carbon-aware optimal energy management.

{\rv It should be noted that the application of SNN features the following two advantages in both CEF learning and carbon-electricity coordinated optimization stages: i) it can learn the complex CEF relationship with combinatorial bilinear terms for high accuracy, compared to the conventional linear approximation; ii) it can achieve fewer variables and constraints for efficient coordinated optimization, compared to the full-connected neural networks. Combinatorial bilinear terms in the CEF model lead to its high non-convexity, which cannot be convexified under current techniques.}

The rest of this paper is organized as follows. Section \ref{sec: prob form} presents the carbon emission analysis in the power allocation process. Section \ref{sec: method} proposes the optimization with the constraint learning framework and corresponding compact constraint learning approach. Section \ref{sec: model} presents the data-driven carbon-aware energy management model based on the proposed constraint learning approach. Section \ref{sec: case} performs the case study to verify the efficiency of the proposed framework, approach, and model. Section \ref{sec: conclusion} concludes this paper.

\section{Carbon Emission Analysis in Power allocation}\label{sec: prob form}

This section firstly analyzes the carbon emission from the generation, network, load, and storage sides based on their different characteristics coupled by the CEF and power flow models, then proposes the CEF model with energy storage, calculates the CEF in the power system, and analyzes the mapping relationships from power injection to power flow to CEF to bus carbon emission for later proposed constraint learning approach. 

\subsection{CEF Analysis}

The CEF model formulates the carbon emission by the CEF rate and CEF intensity, according to Ref.~\cite{Kang2015}. The CEF rate denotes the quantity of CEF through some branch or node per unit of time with the unit of tCO$_2$/h, describing the ``speed'' of CEF. CEF intensity denotes the ratio of the CEF rate to corresponding active power flow for branches and power injection for nodes with the unit of tCO$_2$/kWh, describing the relationship between the CEF and power flow. The CEF analysis is based on the above CEF rate, CEF intensity, and accompanying power flow analysis for generation, network, bus, and storage sides.

\subsubsection{Generation-Network-Bus Side}

Several basic metrics are defined for later illustration, based on Ref.~\cite{Kang2015}. Generation carbon intensity (GCI) analyzes the carbon emission with per-unit energy from generators on the generation side. Branch carbon intensity (BCI) describes the carbon emission with per-unit energy flow along the transmission line. Node carbon intensity (NCI) is the average carbon emission of each bus per unit injected power flow, where different branch carbon emissions aggregate. Based on CEF intensity, the corresponding CEF rates can be calculated by multiplying the CEF intensity with the accompanied active power flow or injection.

\subsubsection{Storage Side}

Energy storage (ES) can alternate between power generation and load demand by discharging and charging. In the process, ES stores the carbon emission virtually and transfers the carbon emission temporally. We denote the energy storage carbon intensity (ESCI) as $e^{es}_{i, t}$. Then according to Ref.~\cite{Wang2022}, company by the charging and discharging, the ES carbon emission is formulated in \eqref{eq: es flow}.

\begin{subequations}
  \allowdisplaybreaks
  \begin{align}
    E^{es}_{i,t} &= e^{es}_{i, t-1} \psi_{i, t-1} + p^{es, cha}_{i, t} e^{\rv es}_{j,t} \Delta t - p^{es, dis}_{i, t} e^{es}_{i,t} \Delta t \label{eq: es flow a} \\
    E^{es}_{i,t} &= e^{es}_{i, t} \psi_{i, t} \label{eq: es flow b}
  \end{align}
  \label{eq: es flow}
\end{subequations}
where $E^{es}_{i,t}$ is the virtual carbon emission volume corresponding to the stored energy $\psi_{i, t}$; $e_{j,t}$ is the NCI where ES is located; $p^{es, cha}_{i,t}$ and $p^{es, dis}_{i,t}$ are the charging and discharging power of ES $i$ in time $t$. 

Then we replace the $E^{es}_{i,t}$ of \eqref{eq: es flow a} with \eqref{eq: es flow b} and consider the different temporal relationships of ESCI under the charging and discharging states to formulate \eqref{eq: es flow detail}.

\begin{subequations}
  \allowdisplaybreaks
  \begin{align}
  e^{es, dis}_{i, t} \psi_{i, t} &= e^{es}_{i, t-1} \psi_{i, t-1} - p^{es, dis}_{i, t} e_{j,t} \Delta t, \text{for } \mu^{dis}_{i, t} = 1 \label{eq: es flow detail a} \\
  e^{es, cha}_{i, t} \psi_{i, t} &= e^{es}_{i, t-1} \psi_{i, t-1} + p^{es, cha}_{i, t} e^{es}_{i, t} \Delta t, \text{for } \mu^{cha}_{i, t} = 1 \label{eq: es flow detail b}
  \end{align}
  \label{eq: es flow detail}
\end{subequations}

The above relationship is non-convex with bilinear terms in the equations. At the charging state of \eqref{eq: es flow detail a}, the ESCI $e^{es, dis}_{i,t}$ of ES $i$ at time $t$ is determined by 1) the last time ES energy and ESCI; 2) the current ES energy, discharging power, and connected BCI, denoted by \eqref{eq: es flow NN a}, and the corresponding $e^{es, ch}_{i,t}$ equals to 0. At the discharging state, the ESCI $e^{es, dis}_{i,t}$ is determined by: 1) the last time ES energy and ESCI; 2) the current ES energy and discharging power, denoted by \eqref{eq: es flow NN b}, and the corresponding $e^{es, dis}_{i,t}$ equals to 0. We utilize $f^{dis}_{\theta} (\cdot)$ and $f^{ch}_{\theta}$ parameterized by $\theta$ to formulate the above relationship in \eqref{eq: es flow NN}.

\begin{subequations}
  \allowdisplaybreaks
  \begin{align}
    e^{es,dis}_{i,t} &= f^{dis}_{\theta} (\psi_{i, t}, \psi_{i, t-1}, p^{es, dis}, e^{es}_{i, t-1}, e_{j, t}) \label{eq: es flow NN a}\\ 
    e^{es,cha}_{i,t} &= f^{cha}_{\theta} (\psi_{i, t}, \psi_{i, t-1}, p^{es, cha}, e^{es}_{i, t-1}) \label{eq: es flow NN b}
  \end{align}
  \label{eq: es flow NN}
\end{subequations}
where $e^{es,dis}_{i,t}$ and $e^{es,cha}_{i,t}$ are the CEF intensity of ES $i$ in time $t$ during the discharging and charging states.

ESCI is the sum of ESCI at discharging and charging states in \eqref{eq: es flow constr}. The big-M method can avoid simultaneously charging and discharging ESCI.

\begin{subequations}
  \allowdisplaybreaks
  \begin{align}
    & e^{es}_{i,t} =  e^{es, dis}_{i,t} + e^{cha}_{i,t} \\ 
    & (1-\mu^{cha}_{i,t}) M^{L} \leq e^{cha}_{i,t} - f^{cha}_{\theta} (\cdot) \leq (1 - \mu^{cha}_{i,t}) M^{H} \\
    & \mu^{cha}_{i,t} M^{L} \leq e^{cha}_{i,t} \leq \mu^{cha}_{i,t} M^{H} \\
    & (1-\mu^{dis}_{i,t}) M^{L} \leq e^{dis}_{i,t} - f^{dis}_{\theta} (\cdot) \leq (1-\mu^{dis}_{i,t}) M^{H} \\ 
    & \mu^{dis}_{i,t} M^{L} \leq e^{dis}_{i,t} \leq \mu^{dis}_{i,t} M^{H}
  \end{align}
  \label{eq: es flow constr}
\end{subequations}
where $\mu^{dis}_{i,t}$ and $\mu^{cha}_{i,t}$ are binary indicator variables against simultaneously charging and discharging; $M^{H}$ and $M^L$ are the positive and negative large values for the big-M method to relax bilinear terms with integer variables

\subsection{Carbon Flow Model with Energy Storage}

CEF calculation is based on the generation, network, node, and storage side carbon analysis. Based on the Ref. \cite{Cheng2019}, the node carbon emission intensity (NCI) with ES is further derived as the weighted average of injected branch flow emission (BCI) and generation (GCI), as illustrated in \eqref{eq: cef node} and \eqref{eq: cef branch}. 

\begin{eqnarray}
  {e}_{i,t} = \frac{\sum_{i\in\mathcal{N}^G_i} {e}^{G}_i {p}^G_{i,t} + \sum_{j\in\mathcal{N}^{l+}_i} {\rho}_{ij,t}| {p}_{ij,t}| + \sum_{i\in\mathcal{N}^{ES}_{i}} p^{dis}_{i,t}e^{dis}_{i,t}}
    {\sum_{i\in\mathcal{N}^G_i} {p}^G_{i,t} + \sum_{j\in\mathcal{N}^{l+}_i} |{p}_{ij,t}| + \sum_{i\in\mathcal{N}^{ES}_{i}} p^{dis}_{i,t}} \label{eq: cef node a}
  \label{eq: cef node}
\end{eqnarray}
where $\mathcal{N}^G_i$, $\mathcal{N}^{l+}_i$, and $\mathcal{N}^{ES}_{i}$ are the sets of generation, flow-in branches, and energy storage with bus $i$; $e_{i,t}$ is the nodal carbon emission intensity for bus $i$ in time $t$.

\begin{eqnarray}
  {\rho}_{ij,t} = \left\{
    \begin{matrix}
      {e}_{i,t} & \quad {P}_{ij,t} \geq 0 \\ 
      {e}_{j,t} & \quad {P}_{ij,t} < 0 
    \end{matrix} 
  \right. \label{eq: cef node b}
  \label{eq: cef branch}
\end{eqnarray}
where $p_{ij,t}$ and $\rho_{ij,t}$ are the active power flow and corresponding BCI from bus $i$ to bus $j$ in time $t$. BCI calculation follows the proportional sharing principle from Ref.~\cite{Kang2015}.

\subsection{Carbon Flow Calculation}

Calculating the carbon flow based on \eqref{eq: cef node} is ineffective. {\rv Its ineffectiveness lies in the high complex nonlinear and non-convex relationship between power injection and carbon emission, which cannot be embedded in the conventional energy management model to achieve the {effective} carbon-electricity coordinated optimization.} Considering the relationship between the NCI and GCI, the numerator of \eqref{eq: cef node a} can be written as:
\begin{eqnarray}
  \begin{aligned}
    \sum_{i\in\mathcal{N}^G_i}{e}^{G}_i {p}^G_{i,t} + \sum_{j\in\mathcal{N}^{l}_i}{\rho}_{ij,t}|{p}_{ij,t}| + \sum_{i\in\mathcal{N}^{ES}_{i}} p^{dis}_{i,t}e^{dis}_{i,t}  \\ 
    = \eta^{(i)}_{N}(\pmb{P}^T_B \pmb{e}_N + \pmb{P}^T_G \pmb{e}_G + \pmb{P}^T_{dis} \pmb{e}_{dis})
  \end{aligned}
  \label{eq: cef numerator}
\end{eqnarray}
where {\rv $\eta^{(i)}_{N}$ refers to an $N$ dimension row vector, where $ith$ element of $\eta^{(i)}_{N}$ is one and the rest elements are zero;} $\pmb{e}_{N}$ is the node carbon emission intensity; $\pmb{P}_B$ is the branch outflow power distribution matrix. $\pmb{P}_B$ describes the outflow power distribution in the power system. For the element of $\pmb{P}_B$, if the active power $p$ flows from bus $i$ to bus $j$, then $P_{B, ij} = p$, and $P_{B, ji} = 0$. {\rv We note that current convexification techniques may deal with single bilinear term constraints but cannot convexify the combinatorial bilinear terms on the left-hand side of \eqref{eq: cef numerator}.}

Then the denominator of \eqref{eq: cef node a} can be written as:
\begin{eqnarray}
  \sum_{i\in\mathcal{N}^G_i} {p}^G_{i,t} + \sum_{j\in\mathcal{N}^{l}_i} |{p}_{ij,t}| + \sum_{i\in\mathcal{N}^{ES}_{i}} p^{dis}_{i,t} 
  = \eta^{(i)}_{N} \pmb{P}_N (\eta^{(i)}_{N})^T
  \label{eq: cef denominator}
\end{eqnarray}
where $\pmb{P}_N$ is the diagonal matrix with the element as $\pmb{P}_{Nii} = \sum_{i\in\mathcal{N}^G_i} {P}^G_{i,t} + \sum_{j\in\mathcal{N}^{l}_i} |{P}_{ij,t}| + \sum_{i\in\mathcal{N}^{ES}_{i}} p^{dis}_{i,t}$ and so the $\pmb{P}_N$ can be calculated by \eqref{eq: p_n cal}.

\begin{eqnarray}
  \pmb{P}_N = diag\{\zeta_{N+K+M} [\pmb{P}_B^T \quad \pmb{P}_G \quad \pmb{P}_{dis}]^T \}.
  \label{eq: p_n cal}
\end{eqnarray}

Then based on \eqref{eq: cef numerator} and \eqref{eq: cef denominator}, the CEF calculation with ES model is reformulated in \eqref{eq: cef general}.
\begin{subequations}
  \allowdisplaybreaks
  \begin{align}
    \pmb{P}_N \pmb{e}_N &= \pmb{P}^T_B \pmb{e}_N + \pmb{P}^T_G \pmb{e}_G + \pmb{P}^T_{dis} \pmb{e}_{dis} \label{eq: cef general a}\\
    \pmb{e}_N &= (\pmb{P}_N - \pmb{P}^T_B)^{-1} (\pmb{P}^T_G \pmb{e}_G+ \pmb{P}^T_{dis} \pmb{e}_{dis}) \label{eq: cef general b} \\
    \pmb{R}_L &= \pmb{P}_L \circ \pmb{e}_N \label{eq: cef general c}
  \end{align}
  \label{eq: cef general}
\end{subequations}
where $\circ$ is the element-wise multiplication operator; $\pmb{P}_L$ and $\pmb{R}_L$ are the load {\rv demand} and corresponding CEF rate. 

\subsection{Mapping Relationship Analysis}

Eq.~\eqref{eq: cef general b} mapping from the vector $\pmb{M}$ composed of ($\pmb{P}_N$, $\pmb{P}_B$, $\pmb{P}_G$, $\pmb{e}_G$, $\pmb{P}_{dis}$, $\pmb{e}_{dis}$) to the nodal CEF rate of $\pmb{R}_L$, and $\pmb{M}$ is determined by the power flow from the node power injection $\pmb{P}_{inj} = (\pmb{P}_G, \pmb{P}_D)$ utilizing the linearized power flow model. So we can formulate the following mapping relationship by $g_{pf}(\cdot)$ and $f_{cf}(\cdot)$ in \eqref{eq: semi map}.

\begin{subequations}
  \allowdisplaybreaks
  \begin{align}
    & \pmb{P}_{inj}\rightarrow \pmb{M} \rightarrow \pmb{R}_L \label{eq: semi map a}\\
    & g_{pf}: \pmb{P}_{inj}\rightarrow \pmb{M} \quad \text{Mapping via power flow \eqref{eq: constr acpf}} \label{eq: semi map b} \\ 
    & f_{cf}: \pmb{M} \rightarrow \pmb{R}_L \quad \text{Mapping via CEF \eqref{eq: cef general}} \label{eq: semi map c}
  \end{align}
  \label{eq: semi map}
\end{subequations}
where the mapping of \eqref{eq: semi map b} is a linear transformation. In contrast, the mapping of \eqref{eq: semi map c} is a complex transformation, which is highly non-convex due to the matrix inverse operation and bilinear terms. So the final mapping of \eqref{eq: map} is also highly non-convex, which cannot be encoded in the optimization model as tractable constraints.
\begin{eqnarray}
  \pmb{R}_L = f_{cf}(g_{pf}(\pmb{P}_{inj})).
  \label{eq: map}
\end{eqnarray}

So we take advantage of the highly representational machine learning techniques to approximate the above composite mapping $f_{cf}(g_{pf}(\cdot))$ by $f_{\theta}(\cdot)$ parameterized with $\theta$ through the proposed constraint learning approach in section \ref{sec: method}.
\begin{eqnarray}
  \hat{\pmb{R}}_L = f_{\theta}(\pmb{P}_{inj}).
  \label{eq: map general}
\end{eqnarray}

{\rv 
\begin{remark}
  Though only the mapping $f_{cf}(\cdot)$ is highly non-convex, we propose to learn the composite mapping of $f_{cf}(g_{pf}(\cdot))$. The main reason behind learning $f_{cf}(g_{pf}(\cdot))$ rather than $f_{cf}(\cdot)$ lies in $\pmb{M}$ features more dimensions than $\pmb{P}_{inj}$. It will pose two problems: 1) sampling data from $\pmb{M}$ is more complicated than sampling data from $\pmb{P}_{inj}$; 2) the transformed SNNs will lead to more variables and constraints for efficient coordinated optimization. 
\end{remark}
}

\begin{remark}
  For some topology power network structures or operating states, such as an isolated bus, there might be zero diagonal elements of $\pmb{P}_N$, leading to its invertibility in \eqref{eq: cef general}. It can be fixed by eliminating the relevant buses and matrices.
\end{remark}

\section{Methodology} \label{sec: method}

This section introduces the optimization with the constraint learning framework for carbon emission management. And then, it proposes the compact constraint learning approach with a sparse neural network design for approximating the complex CEF model of \eqref{eq: es flow constr} and \eqref{eq: map general} and transforming them into {\rv a series of} MILP constraints.

\subsection{Optimization With Constraint Learning Framework under Data-driven Lens}

Practical optimization problems often contain constraints with intractable forms, e.g., the complex CEF models of \eqref{eq: es flow constr} and \eqref{eq: map general}. However, based on the supportive data corresponding to such constraints, ML can utilize these data to learn the above complex constraints, denoted by constraint learning, CL. Then the trained ML models can be transformed into tractable constraints and embedded into the optimization models. So the above two stages are called optimization with constraint learning (OCL). The final ML-encoded tractable model is called the data-driven optimization model. It should be noted that the OCL framework can apply to objective learning by transforming the objective function into the equivalent constraint via its epigraph \cite{Maragno2021}. 

According to Ref.~\cite{Fajemisin2021}, the OCL framework comprises the following five steps: i) construct the conceptual optimization model for problem formulation; ii) conduct data sampling and processing; iii) construct and train ML models for constraint learning; iv) transform trained ML models into MILP and embed them into the optimization model; v) solve and evaluate the data-driven optimization model. {\rv We note that the proposed compact constraint learning approach leverages i) the high representational capability for the accurate CEF approximation compared with other convexification and linearization approaches ii) and its MILP representability for the final MILP-based data-driven carbon-aware energy management model in \eqref{eq: ems} compared to the nonlinear heuristic algorithm.}

Under the above OCL framework, we propose the SNN to learn the complex CEF model of \eqref{eq: es flow constr} and \eqref{eq: map general}, transform them into equivalent MILP constraints, and encode them in the data-driven carbon-aware energy management model, detailed in the following parts. {\rv Compared with the conventional ML model, the main advantage of the proposed compact constraint learning with SNN lies in its high representational capability for the accurate CEF learning and the MILP representability for the efficient carbon-electricity coordinated optimization.} 

\subsection{Compact Constraint Learning Approach with Sparse Neural Networks for Carbon Emission Flow}

\begin{figure}[ht]
  \centering
  \includegraphics[scale=1]{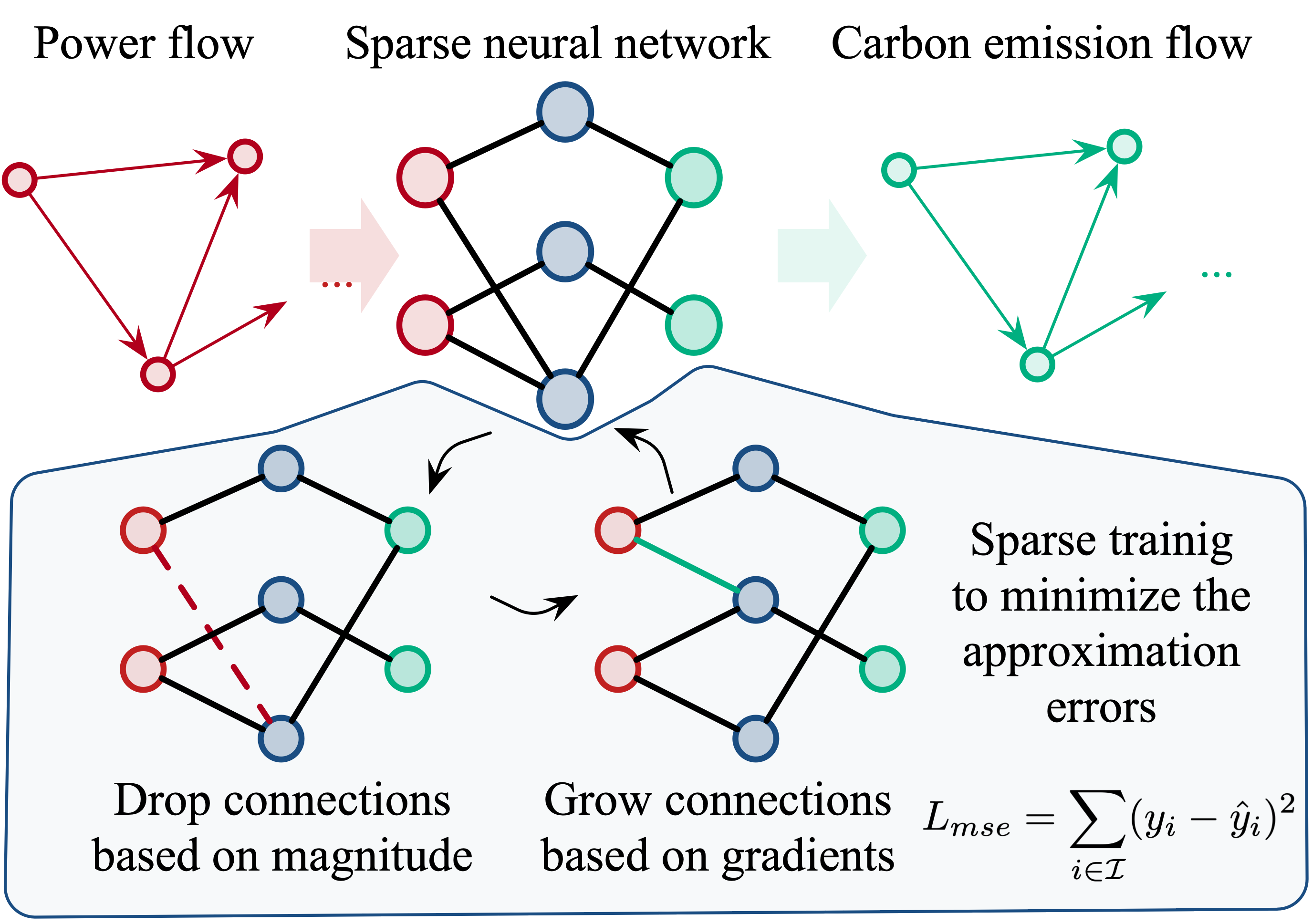}
  \caption{Compact constraint learning with sparse neural networks.}
  \label{fig: cl method}
\end{figure}

Based on the OCL framework, we propose the compact constraint learning approach with sparse neural networks for mapping power flow to the corresponding CEF, as illustrated in Fig.~\ref{fig: cl method}. {\rv Specifically, our proposed compact constraint learning is composed of two stages: 1) it utilizes the sparse neural networks (SNNs) to learn the complex relationship between the bus power injection and bus carbon emission via \eqref{eq: es flow constr} and \eqref{eq: map general}, which refers to the ``learning'' of the compact constraint learning; 2) the trained SNNs with fewer parameters are transformed into a series of  fewer variables and mixed-integer linear constraints via \eqref{eq: cef snn} and embedded into energy management model to achieve efficient carbon-electricity coordinated optimization, where the fewer variables and mixed-integer linear constraints refer to the ``compact'' and ``constraint'' of the compact constraint learning.}

SNN features comparable approximation capability with dense NN but fewer parameters than dense NN, which can facilitate the inference floating-point operations and require less storage space. The architecture of SNN is the same as dense NN.

\subsubsection{Neural Networks Architecture}

So this paper considers the general form of an $L$-layer feedforward neural network (NN) with rectified linear activation function (ReLU) as the activation function mapping from $x$ to $y$ as $y=f_{NN}(x)$, where $f_{NN}(\cdot)$ is composed of stacked linear transformation and ReLU activation {\rv functions}, in \eqref{eq: nn form}.

\begin{subequations}
  \allowdisplaybreaks
  \begin{align}
    &v^l_i = \text{ReLU} (\theta^l_{i0} + \sum_{j\in\mathcal{N}^{l-1}} \theta^l_{ij} v^{l-1}_j), l\in [L], i\in[N_l] \label{eq: nn form a} \\
    &\text{where  } {y} = v^L = \theta^L_{0} + \sum_{j\in\mathcal{N}^{L-1}} \theta^L_{j} v^{L-1}_j, x=  v^0 \label{eq: nn form b}
  \end{align}
  \label{eq: nn form}
\end{subequations}
where $[L]$ and $[N_l]$ are the sets of NN layers and the neurons in layer $l$, respectively. {\rv Compared with other neural networks, the $L$-layer feedforward NN with ReLU features high representational capacity and computation efficiency for learning the CEF model accurately; the ReLU function can be transformed into a series of mixed integer linear programming (MILP) constraints via the big-M method in \eqref{eq: big-m}, which is suitable for the carbon-electricity coordinated optimization.}

\subsubsection{Model Training}

{\rv The objective of model training is to minimize the \textit{mean-square-error} (MSE) between the predicted value $\hat{y}$ and the actual value ${y}$ in \eqref{eq: nn obj} by optimizing the NN parameters $\theta$ in a stochastic gradient descent way. 

{\rv
\begin{subequations}
  \allowdisplaybreaks
  \begin{align}
      \min \quad & L_{mse} = \sum_{i \in \mathcal{I}} (y_i - \hat{y}_i)^2 \\ 
      \text{s.t. } &\hat{y}_i = f_{\theta}(x) 
  \end{align}
  \label{eq: nn obj}
\end{subequations}
}
For our CEF learning, it belongs to the regression problem, so we leverage the \emph{mean-square-error} to evaluate the prediction error and generate the corresponding gradients $\partial L_{mse} / \partial \hat{y}$ to feed into neural networks for CEF learning, with the help of advanced optimization algorithms, e.g., Adam, RMSProp.}

\subsubsection{Sparse Training}

In addition to conventional NN training, we drop and activate NN inner connections iteratively to alleviate the calculation burden of dense full-connected NN for training SNN. As shown in Fig.~\ref{fig: cl method}, the key to the above sparse training lies in updating neural network topology, which drops connections based on the parameter magnitude criterion and activates connections based on the infrequent gradient criterion from the lottery ticket hypothesis \cite{Frankle2019}. So the main parts of sparse training are composed of: i) sparse distributing, ii) update schedule, iii) connection dropping, and iv) connection growing.

\paragraph*{i) Sparse Distributing}

Sparsity $s^l \in (0, 1)$ is defined as the dropping fraction of the $l^{th}$ layer. Then we initialize the first layer as the dense layer and the rest layers as the sparse layers with predefined $s^l$ for the $l^{th}$ layer sparse training in a uniform way. 

\paragraph*{ii) Update Schedule}

After parameter initialization, the update schedule drops a fraction of connections based on NN parameter magnitude and activates new ones based on NN parameter gradient magnitude at $\Delta T$ interval. So the updating fraction function by considering decaying is defined in \eqref{eq: frac up}.
\begin{eqnarray}
  \begin{aligned}
    f_{decay}(t; \alpha, T_{end}) = \frac{\alpha}{2} (1+\cos \frac{t\pi}{T_{end}})
  \end{aligned}
  \label{eq: frac up}
\end{eqnarray}
where $T_{end}$ is the number of iterations for sparse training, and $\alpha$ is the initial drop fraction.

\paragraph*{iii) Connection Dropping}
{\rv This procedure focuses on dropping the unnecessary connections of the neural networks by selecting small weight parameter absolute values for each layer, indicating the weak impact connection between neurons in the forward propagation. Concretely,} every $\Delta T$ steps, we utilize $ArgTopK(v,k)$ algorithm to select the top-$k$ maximum element indices of vector $v$ and drop the connection by $ArgTopK(-|\theta^l|, f_{decay}(t;\alpha, T_{end})(1-s^l)N^l)$.

\paragraph*{iv) Connection Growing}
{\rv This procedure focuses on growing the new connections by selecting the inactive connections with high magnitude gradients, indicating the strong impact connection between neurons in the backward propagation. Concretely, }after dropping, we further re-activate $k$ connections with the top-$k$ high magnitude gradients by $ArgTopK(\nabla_{\theta^l}L_t , f_{decay}(t;\alpha, T_{end})(1-s^l)N^l)$. Newly activated connections are \textit{initialized to zero} and will not influence the output of the NN.

The above four procedures are applied to each layer of NN sequentially, so the dense layers can be sparsified by selecting the top connections. {\rv The dropping and growing of connections are implemented in the heuristic way and lack the theoretical analysis, where the NN sparsity design is still an open question \cite{Frankle2019}.}

\subsubsection{Sparse SGD Algorithm}

Based on the MSE of \eqref{eq: nn obj} and sparse training designs, we propose the sparse stochastic gradient descent (S-SGD) algorithm for training SNN in algorithm~\ref{algo: cs-sgd}.

\begin{algorithm}[ht]
  {{
    \begin{algorithmic}[1]
      \STATE \textbf{Input:} Prediction model: $f_{\theta}(\cdot)$, sparse distributing: $\mathbb{S}=\{s^1, .., s^L\}$, updating scheduling: $\Delta T$, $T_{end}$, $\alpha$, $f_{decay}$;
      \FOR {$t=1,...,T$}
      \STATE Calculate MSE loss $L_{mse,t} = \sum L_{mse}(f_{\theta}(x_i), y_i)$;
      \IF {$t \bmod \Delta T$ is $0$ and $t \leq T_{end}$}
      \FOR {$l= 1,...,L$}
      \STATE Calculate dropping and activating number \\ $k = f_{decay}(t;\alpha, T_{end})(1-s^l)N^l$;
      \STATE Drop $ArgTopK(-|\theta^l|, k)$;
      \STATE Activate $ArgTopK(\nabla_{\theta^l}L_{mse,t} , k)$;
      \ENDFOR
      \ELSE
      \STATE $\theta \leftarrow \theta - \alpha \Delta_{\theta} L_{mse, t}$
      \ENDIF
      \ENDFOR
      \STATE \textbf{Output:} Trained sparse neural network $f_{\theta}(\cdot) $ parameterized by sparse $\theta$.
    \end{algorithmic}}
  }
  \caption{Sparse Stochastic Gradient Descent Training Algorithm.}
  \label{algo: cs-sgd}
\end{algorithm}

\subsubsection{Exact Equivalent Transformation}

After S-SGD training, SNN can be transformed into the exact equivalent MILP constraints and embedded into the optimization models. Considering the architecture of SNN in \eqref{eq: nn form}, the main nonlinear component lies in the activation function. For each layer of \eqref{eq: nn form a}, the ReLU operator, $\text{ReLU}(x):=\max\{0, x\}$, can be encoded as linear constraints by the big-M method from Ref.~\cite{Maragno2021}, as follows:
\begin{subequations}
  \allowdisplaybreaks
  \begin{align}
    v & \geq x, \\
    v & \leq x - M^{L}(1-t), \\
    v & \leq M^{U} t, \\
    v & \geq 0, \quad t \in \{0, 1\}
  \end{align}
  \label{eq: big-m}
\end{subequations}
{\rv where  $M^{L} < 0$ is a lower bound on all possible values of $x$, and $M^{U} > 0$ is an upper bound. These two auxiliary variables are utilized to restrict the $v$ to $x$ when $x>=0$ and confine the $v$ to 0 when $x<0$, which transforms the $\text{ReLU}(x)$ function into a series of equivalent mixed integer linear constraints.}

{\rv  
\begin{remark}
  Overfitting could threaten the prediction generality of the machine learning models on unseen scenarios, where training and testing sets are different. Our energy management focuses on the optimal operation under normal scenarios. For our CEF learning problem with enough samples for possible scenarios, overfitting is adopted to approximate the complex non-linear CEF model as accurate as possible for all sampled scenarios.
\end{remark}
}

\subsubsection{SNN Embedded Carbon Emission Flow}

Based on the equivalent transformation and sparse training, for each period, we formulate the SNN embedded CEF model in the MILP form by replacing the $\text{ReLU}(x)$ of \eqref{eq: nn form} by \eqref{eq: big-m} for approximating the \eqref{eq: map} in the following \eqref{eq: cef snn}. We replace the $\pmb{P}_{inj}$ with $\pmb{x}$ and $\hat{\pmb{R}}_{L}$ with $\pmb{\hat{y}}$ in a general form.
\begin{subequations}
  \allowdisplaybreaks
  \begin{align}
    v^{1}_{i} &= \theta^1_{i0} + \sum_{j\in\mathcal{X}} \theta^l_{ij} x_j, & & i \in [N_1] \\ 
    \hat{v}^l_i &= \theta^l_{i0} + \sum_{j\in\mathcal{N}^{l-1}} \theta^l_{ij} v^{l-1}_j, & &l\in [L], i\in[N_l] \\
    {v}^l_i &\geq \hat{v}^l_i, & &l\in [L], i\in[N_l] \\ 
    {v}^l_i &\leq \hat{v}^l_i - M^{L}_{l,i} (1-t_{l,i}), & &l\in [L], i\in[N_l] \\ 
    {v}^l_i &\leq M^{H}_{l,i} t_{l,i}, & &l\in [L], i\in[N_l] \\ 
    {v}^l_i &\geq 0, \quad t_{l,i} \in \{0, 1\}, & &l\in [L], i\in[N_l] \\ 
    \hat{y}_m &= \theta^L_{i0} + \sum_{j\in\mathcal{N}^{L-1}} \theta^L_{ij} v^{L-1}_j, & &m\in[N_M], i\in[N_L].
  \end{align}
  \label{eq: cef snn}
\end{subequations}

The corresponding training and transformation for \eqref{eq: es flow NN} can be derived similarly. 

So the intractable complex CEF model of \eqref{eq: es flow NN} and \eqref{eq: cef snn} are approximated via SNNs of \eqref{eq: nn form}, trained via S-SGD of algorithm \ref{algo: cs-sgd}, and encoded as exact equivalent MILP constraints via \eqref{eq: cef snn}.

\section{Data-driven Carbon-aware Energy Management Model} \label{sec: model}

Based on the compact constraint learning with SNN and its equivalent MILP transformation, we design the blocked carbon price mechanism in light of the "high emission with high price" principle and formulate the data-driven carbon-aware energy management (CA-EM) model to incentivize the carbon reduction potential on the demand side.

\subsection{Carbon-aware Optimization with Constraint Learning Framework}

\begin{figure}[ht]
  \centering
  \includegraphics[scale=1.2]{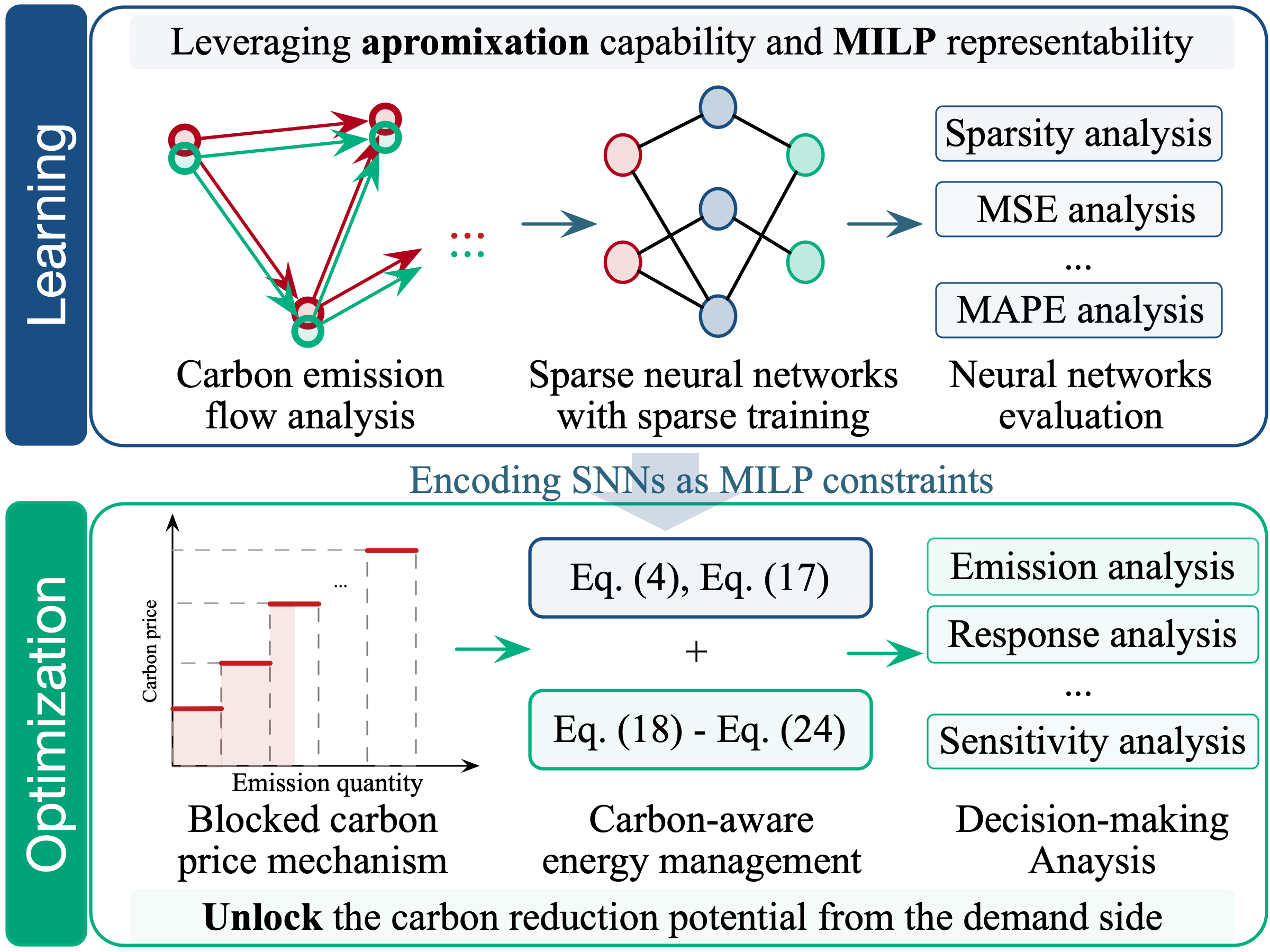}
  \caption{Carbon-aware optimization with constraint learning framework.}
  \label{fig: ocl framework}
\end{figure}

Fig.~\ref{fig: ocl framework} presents the carbon-aware optimization with constraint learning framework for the CA-EM model, which is composed of the upstream constraint learning and downstream carbon-aware optimization. Under the above framework, SNNs learn and transform the CEF models as shown in the top of Fig.~\ref{fig: ocl framework}; then based on the trained SNNs and the blocked carbon price mechanism, CA-EM schedules the system operation in a carbon-aware way as shown in the bottom of Fig.~\ref{fig: ocl framework}. The learning part can leverage the approximation capability and MILP representability of SNN, and the optimization part can unlock the carbon reduction potential from the demand side.

\subsection{Blocked Carbon Price Mechanism}

\begin{figure}[ht]
  \centering
  \subfigure[Blocked carbon price]{\includegraphics[scale=1]{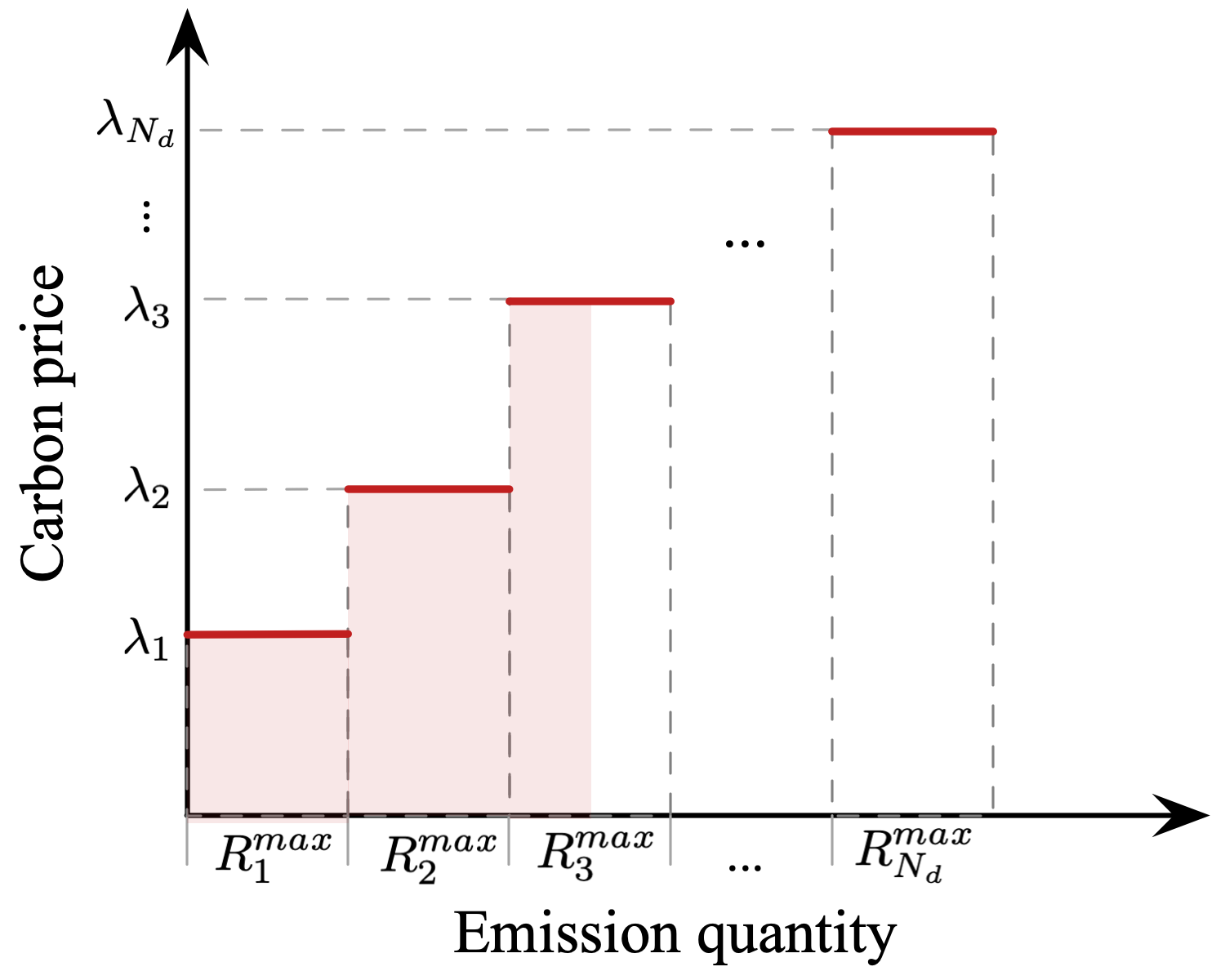}\label{fig: carbon block a}}
  \subfigure[Piece-wise carbon emission cost.]{\includegraphics[scale=1]{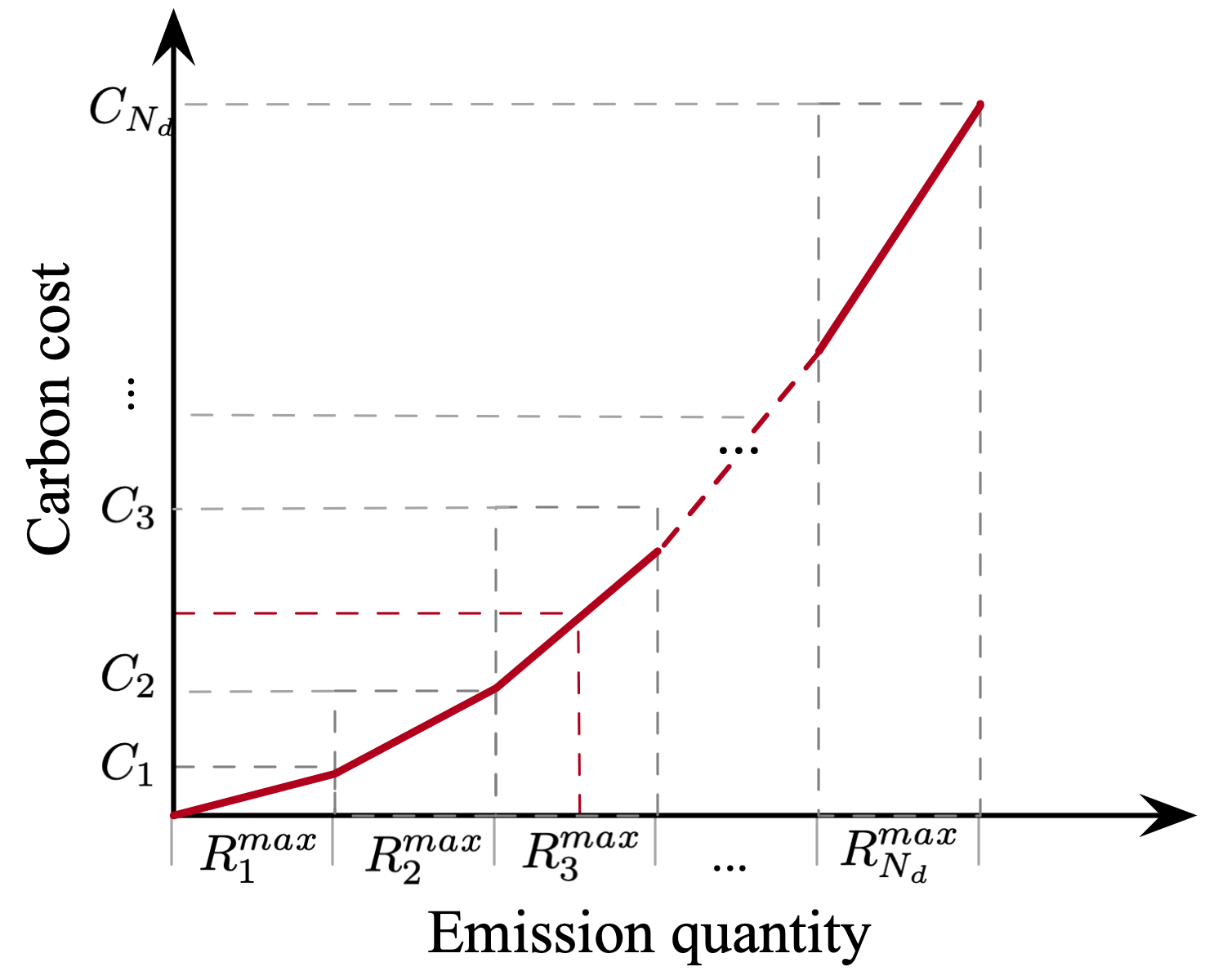}\label{fig: carbon block b}}
  \caption{Blocked carbon price mechanism.}
  \label{fig: carbon block}
\end{figure}

Inspired by the time-of-use pricing strategy \cite{Nguyen2016, Yang2019, Yi2020}, we design a blocked carbon price mechanism. It prices the cost of carbon on the demand side in a blocked way to unlock the carbon reduction potential and incentivize the carbon reduction from the demand side, as shown in Fig.~\ref{fig: carbon block}. {\rv The main motivation behind the proposed mechanism is to penalize high emission demand with high price. And the detailed design refers to the blocked utility function design, where the carbon emission against the energy decarbonization can be treated as the negative utility.} {\rv It should be noted that our proposed blocked carbon price mechanism focuses on unlocking the carbon reduction potential from the demand side, which is distinguished from incentivizing carbon reduction potential from the generation side under the current carbon trading mechanism \cite{Lin2019}.}

The proposed carbon price mechanism separates the price of carbon emission on the demand side into several blocks based on different emission intervals in Fig.~\ref{fig: carbon block a}. So the corresponding carbon emission cost is the piecewise function in terms of emission in Fig.~\ref{fig: carbon block b}.

Mathematically, Eq.~\eqref{eq: carbon pwl} calculates the carbon emission costs for load consumption on the demand side.

\begin{subequations}
  \allowdisplaybreaks
  \begin{align}
    C^{CB}(R_{d,t}) &= \sum_{n=1}^{N_d} \lambda_{n,t} R_{n,t} \Delta T \\ 
    R_{d, t} &= \sum_{n=1}^{N_d} R_{n,t} \\ 
    0 \leq & R_{n,t} \leq R^{max}_{n,t} 
  \end{align}
  \label{eq: carbon pwl}
\end{subequations}
where $R_{n,t}$ is carbon emission interval of block $n$ with unit of tCO$_2$/h., and $p_{n,t}$ carbon price of block $n$ in time $t$ with unit of \$/tCO$_2$. {\rv The above proposed blocked carbon price mechanism can price the carbon emission responsibility in the demand side via and incentivize the carbon reduction potential by demand response via \eqref{eq: ems}.}

\subsection{Carbon-aware Energy Management Model}

Based on the price mechanism of \eqref{eq: carbon pwl} and equivalent MILP SNNs of \eqref{eq: cef snn}, we formulate the final data-driven CA-EM model to maximize the social welfare by subtracting demand utility by the fuel cost, battery, and carbon cost in the demand side.

\subsubsection{Demand Utility Formulation}

The utility function of each user is formulated as follows:
\begin{eqnarray}
  U(P_{d,t}) = \left\{
    \begin{matrix}
      &\beta_{d} (P_{d,t})^2 - \alpha_{d} P_{d,t} & \text{For } P_{d,t}\leq \frac{\beta_{d}}{2 \alpha_{d}} \\ 
      &\frac{\beta_{d}^2}{2 \alpha_{d}} & \text{For } P_{d,t}\geq \frac{\beta_{d}}{2 \alpha_{d}}
    \end{matrix}
  \right. 
  \label{eq: ems utility} 
\end{eqnarray}
where $P_{d,t}$ is the load demand of $d$ in $t$ period; $\alpha_{d}$ and $\beta_{d}$ are corresponding utility coefficient of $d$. Eq.~\eqref{eq: ems utility} can be transformed into a piecewise function based on Ref. \cite{Wang2020}.

\subsubsection{Generators Constraints}

The conventional fuel-based generators power output $p_{g, t}$ are constrained by their minimum/maximum power output ($p^{min}_{g}$/$p^{max}_{g}$) and the ramping constraints of different generators  in \eqref{eq: constr ramp}.

\begin{subequations}
  \allowdisplaybreaks
  \begin{align}
    & p^{min}_{g} \leq p_{g, t} \leq p^{max}_{g} & &g\in\mathcal{G}, \quad t = 1, ..., T \\
    & p_{g, t} - p_{g, t-1} \leq \overline{R}_g  & &g\in\mathcal{G} \quad t = 2, ..., T \\ 
    & p_{g, t-1} - p_{g, t} \leq \underline{R}_g & & g\in\mathcal{G} \quad t = 2, ..., T  
  \end{align}
  \label{eq: constr ramp}
\end{subequations}
where $\mathcal{G}$ is the set of generators; $T$ is the scheduling time; $\overline{R}_g$ and $\underline{R}_g$ are the maximum upward and downward ramping power.

\subsubsection{Upward and Downward Reserve for Prediction Errors}

System upward/downward ($r^{+}$/$r^{-}$) reserve can balance the photovoltaic (PV), wind power (WP), and load day-ahead prediction errors in energy management from the generator reserve $r^{+}_{g,t}$/$r^{-}_{g,t}$ in \eqref{eq: constr up-down}.

\begin{subequations}
  \begin{align}
    & 0 \leq r^{+}_{g,t} \leq \overline{r}, \quad 0 \leq r^{-}_{g,t} \leq \overline{r} \\ 
    & p^{min}_{g} \leq p_{g,t} - r^{-}_{g,t}, \quad p_{g,t} + r^{+}_{g,t} \leq p^{max}_g \\ 
    & \sum_{g\in\mathcal{G}} r^{+}_{g,t} \geq r^{+}, \sum_{g\in\mathcal{G}} r^{-}_{g,t} \geq r^{-}.
  \end{align}
  \label{eq: constr up-down}
\end{subequations}

\subsubsection{Battery Operation Cost}

The battery operation cost is composed of the charging/discharging cost and battery degradation cost based on Ref.~\cite{Tran2013}.
\begin{eqnarray}
  C_{es,t} = \lambda^{es} (p^{es, cha}_{t} + p^{es, dis}_{i, t}) \Delta t + \lambda^{es} \psi_{es, t} \zeta^{L} \Delta t 
  \label{eq: es cost}
\end{eqnarray}
where $\lambda^{es}$ denotes the ES degradation cost; $\psi_{es, t}$ denotes the power leakage loss.

And the detailed energy storage operational model is presented as follows:
\begin{subequations}
  \allowdisplaybreaks
  \begin{align}
    &\psi_t = \psi_{t-1} + \eta_{ch} p^{es, cha}_{t} \Delta_t - \frac{p^{es, dis}_{t}}{\eta_{dis}} \Delta_t, \quad t\in\mathcal{T} \\
    & 0 \leq p^{es, dis}_{t} \leq \min\{P_{dis}^{max}, \eta_{dis}\frac{\psi_{t-1} - \psi_{min}}{\Delta_t} \}, \quad t\in\mathcal{T} \\
    & 0 \leq p^{es, cha}_{t} \leq \min\{ P_{ch}^{max}, \frac{\psi_{max} - \psi_{t-1}}{\eta_{ch} \Delta_t} \}, \quad t\in\mathcal{T} \\
    & 0 \leq p^{es, dis}_{t} \leq \mu_{dis,t} P^{es, dis}_{max} , \quad t\in\mathcal{T} \\
    & 0 \leq p^{es, cha}_{t} \leq \mu_{ch,t} P^{es, dis}_{min} , \quad t\in\mathcal{T}\\
    & \psi_{min} \leq \psi_t \leq \psi_{max}, \mu_{dis,t} + \mu_{ch,t} \leq 1, \quad t\in\mathcal{T}
  \end{align}
  \label{eq: bess}
\end{subequations}

\subsubsection{AC Power Flow Constraints}

The linearized AC power flow constraints in Ref.~\cite{Trodden2014} are utilized to achieve system power balance in \eqref{eq: constr acpf}.

\begin{subequations}
  \allowdisplaybreaks
  \begin{align}
    & P_{ij} = G_{ij}(2V_{i} - 1) - G_{ij}(V_i + V_j -1) - B_{ij}(\theta_i - \theta_j) \\
    & Q_{ij} = -B_{ij}(2V_{i} - 1) + B_{ij} (V_i + V_j -1) - G_{ij} (\theta_i - \theta_j) \\ 
    & \sum_{j\in\mathcal{G}_{b,i}} P_{ij, t} = \sum_{g\in\mathcal{G}(i)} P_{g, t} + \sum_{w\in\mathcal{W}(i)} P_{w, t} - \sum_{d\in\mathcal{D}(i)}P_{d,t} \\
    & \sum_{j\in\mathcal{G}_{b,i}} Q_{ij} = \sum_{g\in\mathcal{Q}(i)} Q_{g, t} - \sum_{d\in\mathcal{D}(i)}Q_{d,t} \\
    & \underline{V} \leq V_{i,t} \leq \overline{V}
  \end{align}
  \label{eq: constr acpf}
\end{subequations}
where the above symbol illustration can refer to Ref.~\cite{Trodden2014} for detailed delivery.

\subsubsection{Data-driven CA-EM Model}

Based on the above, we formulate the final data-driven carbon-aware energy management model {\rv with the objective of maximizing the social welfare by considering the demand utility, energy storage cost, generation cost, reserve cost, and carbon emission cost} under the blocked carbon price design as follows:
\begin{subequations}
  \allowdisplaybreaks
  \begin{align}
    \max &\sum_{t\in\mathcal{T}} 
    \begin{pmatrix}
      \sum_{d\in\mathcal{D}} U(P_{d,t}) - \sum_{es\in\mathcal{E}} C^{es}_{es, t} \\
      -\sum_{g\in\mathcal{G}}C(P_{g,t}) - \sum_{g\in\mathcal{G}}C(r^{+}_{g,t}, r^{-}_{g,t}) \\
      - \sum_{d\in\mathcal{D}} C^{CB}(R_{d,t})
    \end{pmatrix} \\
    \text{s.t. } 
    & \text{\eqref{eq: cef snn}-\eqref{eq: constr acpf}}.
  \end{align}
  \label{eq: ems}
\end{subequations}
The above data-driven CA-EM is a MILP model and can be solved by the off-the-shelf solver efficiently.

\section{Case Study} \label{sec: case}

We utilize the modified IEEE 30-bus system and 118-bus system to verify the effectiveness of the proposed data-driven model with the proposed compact constraint learning approach. The proposed data-driven model is composed of the learning and optimization stages. So, for each stage, we construct the learning models with the Pytorch package and the optimization models with the Cvxpy package equipped by Gurobi 9.0. All the experiments are implemented in a MacBook Pro laptop with RAM of 16 GB, CPU Intel Core i7 (2.6 GHz). The other detailed systems parameter data are attached in Ref.~\cite{Dataset2022}. 

In terms of organization, this section introduces the basics of data preparation and experiment setting for the constraint learning and carbon-aware optimization in section \ref{sub: dp}, verifies the effectiveness of the constraint learning with sparse neural networks in section \ref{sub: cla}, analyzes the energy management results under the proposed CA-EM model with the blocked carbon price mechanism in section \ref{sub: cda}, and discusses the varying carbon price impact on the CA-EM model in section \ref{sub: sa}.

\subsection{Data Preparation and Experiment Setting} \label{sub: dp}

This part presents the main procedures of data processing and experiment setting of learning/optimization models.

\subsubsection{Data processing}

{\rv
The main procedures of data processing are composed of three main procedures: i) \textit{Data generation}: we generate the essential data by solving energy management scheduling model without carbon price under different load parameters {\rv and set the sampled number as 4000 for the 30-bus system and 10000 for the 118-bus system.} ii) \textit{Feature engineering}: it takes the part of the solved model data as the input vector and calculates corresponding carbon emission as the output vector from \eqref{eq: map general}. iii) \textit{Normalization}: as the units of input vector elements are different, we normalize them into the uniform 0-1 scale for enhancing the approximation accuracy.
}

\subsubsection{Learning Models}

For the learning models, Tab.~\ref{tab: nn param} presents the initialized hyperparameters of multi-layer SNN. {\rv And the parameters of the full connected neural network are set as the same initialized hyperparameters of SNN with the same NN structures to verify the effectiveness of the sparsity design.}
\begin{table}[ht]
  \renewcommand{\arraystretch}{1.3}
  \centering
  \caption{The hyperparameters of neural network models.}
  \begin{tabular}{cc|cc}
    \hline
    Hyperparameter & Value & Hyperparameter & Value \\
    \hline
    Optimizer & Adam & Learning rate & 1e-6 \\ 
    Hidden layers for 30-bus & [1000] & Batch size & 32 \\ 
    Hidden layers for 118-bus & [100, 100] & Activation function & ReLU  \\
    Hidden layers for ES & [50] & Dropout & 0.2 \\
    \hline
  \end{tabular}
  \vspace{1ex}
  
  {\raggedright \rv We note that the above hyperparameters refer to the hyperparameter value of conventional full-connected neural networks and initialized hyperparameter of sparse neural network for sparsity comparison. \par}

  \label{tab: nn param}
\end{table}

\subsubsection{Optimization Models}

For the carbon-aware energy management model, Tab.~\ref{tab: opt param} presents the corresponding key parameters, including the carbon emission intensity of fuel-based generators and related distributed energy resource parameters. And the blocked carbon prices of \eqref{eq: carbon pwl} are set as \$40 /tCO$_2$, \$60 /tCO$_2$, \$80 /tCO$_2$, \$100 /tCO$_2$ with the corresponding interval as 10 tCO$_2$, 10 tCO$_2$/h, 20 tCO$_2$/h, 20 tCO$_2$/h. {\rv It should be noted that the proposed blocked carbon prices are different from the current uniform carbon price, we set the blocked carbon prices by referring to the current uniform carbon price in Ref.~\cite{Lin2019} manually.}

\begin{table}[ht]
  \renewcommand{\arraystretch}{1.3}
  \centering
  \caption{The key parameters of the carbon-aware energy management model.}
  \begin{tabular}{cc|cc}
    \hline
    Hyperparameter & Value & Hyperparameter & Value \\
    \hline
    Coal generator $e$ & 0.875 tCO$_2$/MWh & PV capacity & 50MW \\
    Gas generator $e$ & 0.520 tCO$_2$/MWh & WP capacity & 50MW \\
    Charging depth & 0.5 & Discharging depth & 0.5 \\
    $\eta_{dis}$ & 0.92 & $\eta_{ch}$ & 0.90 \\
    $\Delta_t$ & 1 hour & & \\ 
    \hline
  \end{tabular}
  \label{tab: opt param}
\end{table}

\subsection{Constraint Learning Analysis} \label{sub: cla}

Based on data processing, this part analyzes the approximation performance of the constraint learning for the CEF model of \eqref{eq: cef general} and energy storage model of \eqref{eq: es flow NN} to verify the effectiveness of the sparse neural networks.

\subsubsection{Training Process Analysis}

\begin{figure}[ht]
  \centering
  \subfigure[Training process in the 30-bus system.]{\includegraphics[scale=1]{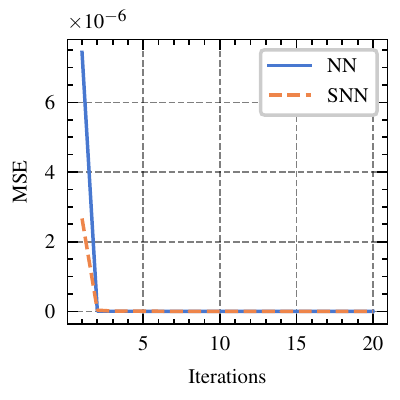}}
  \subfigure[Training process in the 118-bus system.]{\includegraphics[scale=1]{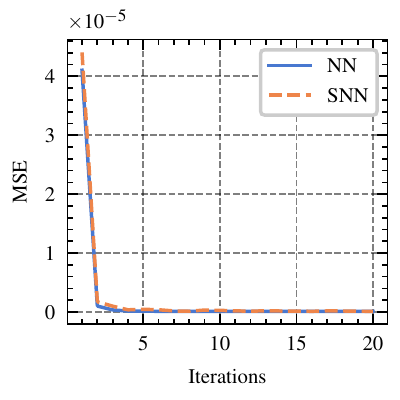}}
  \caption{Training process of different neural networks analysis.}
  \label{fig: training process}
\end{figure}

Fig.~\ref{fig: training process} compares the training process of the conventional dense neural networks (NN) and sparse neural network (SNN) to the mapping relationships of \eqref{eq: es flow NN} and \eqref{eq: cef snn} for the 30-bus and 118-bus systems. NN and SNN converge at about three epochs with similar errors, demonstrating their similar approximation effectiveness.

\subsubsection{Results Analysis}

Then we analyze the performance of the learning models from the perspective of approximation accuracy and sparsity intensity by comparing SNN, NN, and linear regression (LR) models.

\paragraph{Approximation Analysis}

We utilize the mean square error (MSE), root-mean-square error (RMSE), mean absolute percentage error (MAPE) metrics, and $R^2$ to measure the approximation performance of different methods numerically.

Tab.~\ref{tab: c30 pred error} and \ref{tab: c118 pred error} compare the MAPE, MSE, and negative rate of CSNN, C-NN, NN, and LR models in the 30-bus and 118-bus systems. In the below 30-bus system, SNN for \eqref{eq: cef snn} exhibits higher approximation accuracy with 0.517\% MAPE, 5.48$e^{-9}$ MSE, 7.01$e^{-5}$ RMSE, and 0.9991 $R^2$, a litter lower than NN. SNN and NN outperform the LR due to their higher representational capacity with more parameters.
\begin{table}[ht]
  \renewcommand{\arraystretch}{1.3}
  \centering
  \caption{Approximation error comparison of different learning models in the 30-bus system.}
  \begin{tabular}{ccccccc}
    \hline
    \multirow{2}{*}{Metrics} & \multicolumn{3}{c}{CEF} & \multicolumn{3}{c}{ES}\\
    \cline{2-7}
     & SNN & NN & LR & SNN & NN & LR \\
    \hline
    MAPE & 0.52\% & 0.53\% & 0.60\% & 0.34\% & 0.38\% & 0.45\% \\ 
    MSE/$e^{-9}$ & 5.48 & 5.84 & 6.97 & 3.09 & 3.62 & 4.24 \\ 
    RMSE/$e^{-5}$ & 7.01 & 7.23  & 8.35 & 5.02 & 5.58 & 6.08  \\
    $R^2$ & 0.9991 & 0.9990 & 0.9975 & 0.9996 & 0.9994 & 0.9992 \\
    \hline
  \end{tabular}
  \vspace{1ex}
  
  {\raggedright We note that LR denotes the linear regression model. \par}
  \label{tab: c30 pred error}
\end{table}

And in the below 118-bus system, SNN for \eqref{eq: cef snn} also exhibits higher approximation accuracy with 0.656\% MAPE, 5.554$e^{-7}$ MSE, 3.74$e^{-5}$ RMSE, and 0.9938 $R^2$, a litter lower than NN. The approximation error of the 118-bus system is higher than that of the 30-bus system because more buses mean a more complex mapping. 

\begin{table}[ht]
  \renewcommand{\arraystretch}{1.3}
  \centering
  \caption{Approximation error comparison of different learning models in the 118-bus system.}
  \begin{tabular}{ccccccc}
    \hline
    \multirow{2}{*}{Models} & \multicolumn{3}{c}{CEF} & \multicolumn{3}{c}{ES}\\
    \cline{2-7}
     & SNN & NN & LR & SNN & NN & LR \\
    \hline
    MAPE & 0.66\% & 0.67\% & 0.81\% & 0.44\% & 0.45\% & 0.61\%   \\ 
    MSE/$e^{-7}$ & 5.54 & 5.71 & 8.71 & 2.32 & 2.39 & 3.38 \\ 
    RMSE/$e^{-4}$ & 3.74 & 4.0 & 5.29 & 4.89 & 4.92 & 5.82 \\
    $R^2$ & 0.9938 & 0.9936 & 0.9915 & 0.9958 & 0.9954 & 0.9924 \\
    \hline
  \end{tabular}
  \label{tab: c118 pred error}
\end{table}

\paragraph{Sparsity Analysis}

Though NN and SNN feature comparable approximation capability, their sparsity intensities are different due to the additional sparse training of algorithm \ref{algo: cs-sgd}. So Tab.~\ref{tab: c30 param num} and \ref{tab: c118 param num} compares the sparsity intensity of NN and SNN in carbon flow learning of \eqref{eq: cef snn} and charging/discharging learning of \eqref{eq: bess} in 30-bus and 118-bus systems. To evaluate neural network sparsity intensity, we define the sparsity rate as the ratio of the parameter reduction number to the total parameter number. In the 30-bus system, compared with the 0\% sparsity rate of dense NN, the SNN under the sparse training of algorithm \ref{algo: cs-sgd} achieves 41.55\% sparsity rate with 41.55\% fewer parameters in carbon flow learning \eqref{eq: cef snn} and achieves average 40.30\% sparsity rate in ES learning \eqref{eq: es flow NN}.
\begin{table}[ht]
  \renewcommand{\arraystretch}{1.3}
  \centering
  \caption{The parameter number of different learning models for carbon flow approximation in the 30-bus system.}
  \begin{tabular}{cccccccc}
    \hline
    \multirow{2}{*}{Metrics} & \multicolumn{3}{c}{CEF} & \multicolumn{3}{c}{ES}\\
    \cline{2-7}
     & SNN & NN & LR & SNN & NN & LR \\
    \hline
    Number & 39180 & 67030 & 125 & 207 & 122 & 207 \\ 
    Sparsity rate & 41.55\% & 0\% & - & 40.30\% & 0\% & - \\
    \hline
  \end{tabular}
  \label{tab: c30 param num}
\end{table}

In the 118-bus system, compared with the dense NN, the SNN under the sparse training of algorithm \ref{algo: cs-sgd} achieves a 50.96\% sparsity rate with 50.96\% fewer parameters in carbon flow learning \eqref{eq: cef snn} and achieves average 27.26\% sparsity rate in ES learning \eqref{eq: es flow NN}.

\begin{table}[ht]
  \renewcommand{\arraystretch}{1.3}
  \centering
  \caption{The parameter number of different learning models for carbon flow approximation in the 118-bus system.}
  \begin{tabular}{cccccccc}
    \hline
    \multirow{2}{*}{Metrics} & \multicolumn{3}{c}{CEF} & \multicolumn{3}{c}{ES}\\
    \cline{2-7}
     & SNN & NN & LR & SNN & NN & LR \\
    \hline
    Number & 19282 & 39318 & 300 & 409 & 295 & 409 \\ 
    Sparsity rate & 50.96\% & 0\% & - & 27.26\% & 0\% & - \\
    \hline
  \end{tabular}
  \label{tab: c118 param num}
\end{table}

{\rv
\begin{remark}
  Training of SNN and NN indicate similar convergence rate due to the same hyperparameter setting of SNN and NN as shown in Fig.~\ref{fig: training process}, Tab.~\ref{tab: c30 pred error}, and Tab.~\ref{tab: c118 pred error}. The main reason lies in we set the same hyperparameter values for the initialized hyperparameter of SNN and hyperparameter NN to verify the effectiveness of sparsity design, as shown in Tab.~\ref{tab: nn param}. And we further demonstrate the rationality of hyperparameter selection in appendix \ref{apx: nn tuning}.
\end{remark}
}

{\rv
\subsubsection{Discussion and Comparison}

We further compare the training and testing time of the proposed learning method for different cases in Tab.~\ref{tab: time comparison}. Due to sparsity design, SNN achieves less time training time and faster prediction time. For case 30, training SNN requires 36.41\% less time, and prediction through SNN achieves 2.51x faster prediction time for the CEF learning; for case 118, training SNN requires 35.37\% less time, and prediction through SNN achieves 1.80x faster prediction time for the CEF learning. For the ES learning, the training and prediction times feature comparable results.
}

\begin{table}[ht]
  \renewcommand{\arraystretch}{1.3}
  \centering
  \caption{\rv Training and testing time evaluation for carbon emission flow approximation in different cases}
  \begin{tabular}{ccccccc}
    \hline
    \multirow{2}{*}{Models} & \multirow{2}{*}{Types} & \multicolumn{2}{c}{CEF} & \multicolumn{2}{c}{ES}\\
    \cline{3-6}
    & & SNN & NN & SNN & NN \\
    \hline
    \multirow{2}{*}{Case 30} & Training & 482s & 758s & 382s & 451s \\   
    & Testing & 0.00031s & 0.00078s & 0.00028s & 0.00037s \\
    \hline
    \multirow{2}{*}{Case 118} & Training & 3144s & 4865s & 344s & 424s \\   
    & Testing & 0.00061s & 0.0011s & 0.00025s & 0.00032s \\ 
    \hline
  \end{tabular}
  \label{tab: time comparison}
\end{table}

{\rv
Then we further compare the approximation performance of SNN with other common ML models in Tab.~\ref{tab: ml comparison}, including the support vector regression (SVR) and random forest (RF) models from Ref.~\cite{Goodfellow2016}. We note that conventional SVR only applies to single point prediction, so we here adopt the multiple output SVR model in Ref.~\cite{Bao2014} for the model learning. As shown in Tab.~\ref{tab: ml comparison}, the proposed compact constraint learning can achieve higher approximation accuracy in terms of higher $R^{2}$ and lower MSE compared to SVR and RF, verifying its learning effectiveness. For SVR and RF setting, the SVR is with radial basis function kernel, $\gamma$ as 0.3, and $\epsilon$ as 0.01; the RF is with the estimator number as 100, maximum depth as 3, and maximum feature as 1.0.
}
\begin{table}[ht]
  \renewcommand{\arraystretch}{1.3}
  \centering
  \caption{\rv Approximation error comparison with conventional ML models in different cases.}
  \begin{tabular}{ccccccc}
    \hline
     \multirow{2}{*}{Metrics} & \multicolumn{3}{c}{CEF} & \multicolumn{3}{c}{ES}\\
    \cline{2-7}
     & SNN & SVR & RF & SNN & SVR & RF  \\
    \hline
    \emph{Case 30} & - & - & - & - & - & - \\
    MSE/$e^{-9}$ & 5.48 & 7.45 & 9.12 & 3.09 & 4.41 & 5.31 \\   
    $R^2$ & 0.9991 & 0.9971 & 0.9262 & 0.9996 & 0.9992 & 0.9987 \\
    \hline
    \emph{Case 118} & - & - & - & - & - & - \\
    MSE/$e^{-7}$ & 5.54 & 23.58 & 24.62 & 2.39 & 7.23 & 10.31\\
    $R^2$ & 0.9938 & 0.932 & 0.917 & 0.9958 & 0.981 & 0.945 \\
    \hline
  \end{tabular}
  \label{tab: ml comparison}
\end{table}

\subsection{Carbon-aware Energy Management Model Analysis} \label{sub: cda}

Based on the above compact constraint learning approach, this part i) encodes the trained SNN models in the energy management model via \eqref{eq: big-m} to formulate the data-driven CA-EM model of \eqref{eq: ems}, ii) analyzes its low-carbon performance in incentivizing the carbon reduction potential from the demand side, iii) and compares the proposed CA-EM with conventional energy management (EM) model without carbon awareness in the 30-bus and 118-bus systems.

\subsubsection{Case 30-bus Results Analysis}

\begin{figure}[ht]
  \centering
  \includegraphics[scale=0.9]{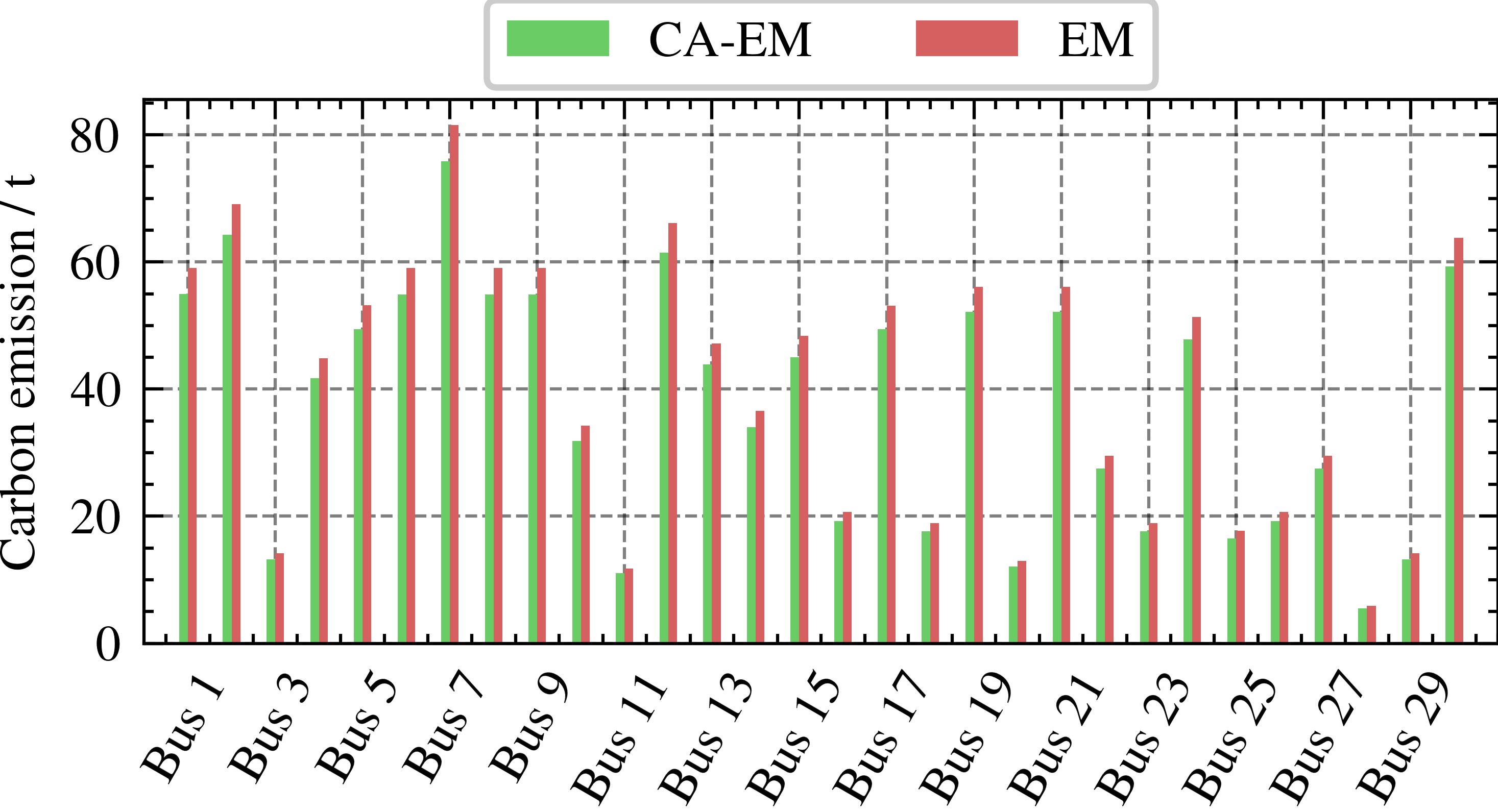}
  \caption{The node carbon emission analysis under different scheduling models in the 30-bus system.}
  \label{fig: c30 node bci}
\end{figure}

For the 30-bus system, Fig.~\ref{fig: c30 node bci} compares the node carbon emission rates of the CA-EM and EM models.  
As shown in Fig.~\ref{fig: c30 node bci}, the proposed CA-EM can reduce each bus carbon emission effectively considering carbon price on the demand side. The emission reductions of the emission-intensive buses are higher with a 5.71 tCO$_2$  reduction in bus seven due to the blocked carbon price mechanism design with a high carbon price penalty on high carbon emission, verifying the proposed blocked price design effectiveness.

\begin{figure}[ht]
  \centering
  \subfigure[Demand response reduction analysis under the CA-EM in the 30-bus system.]{\includegraphics[scale=0.50]{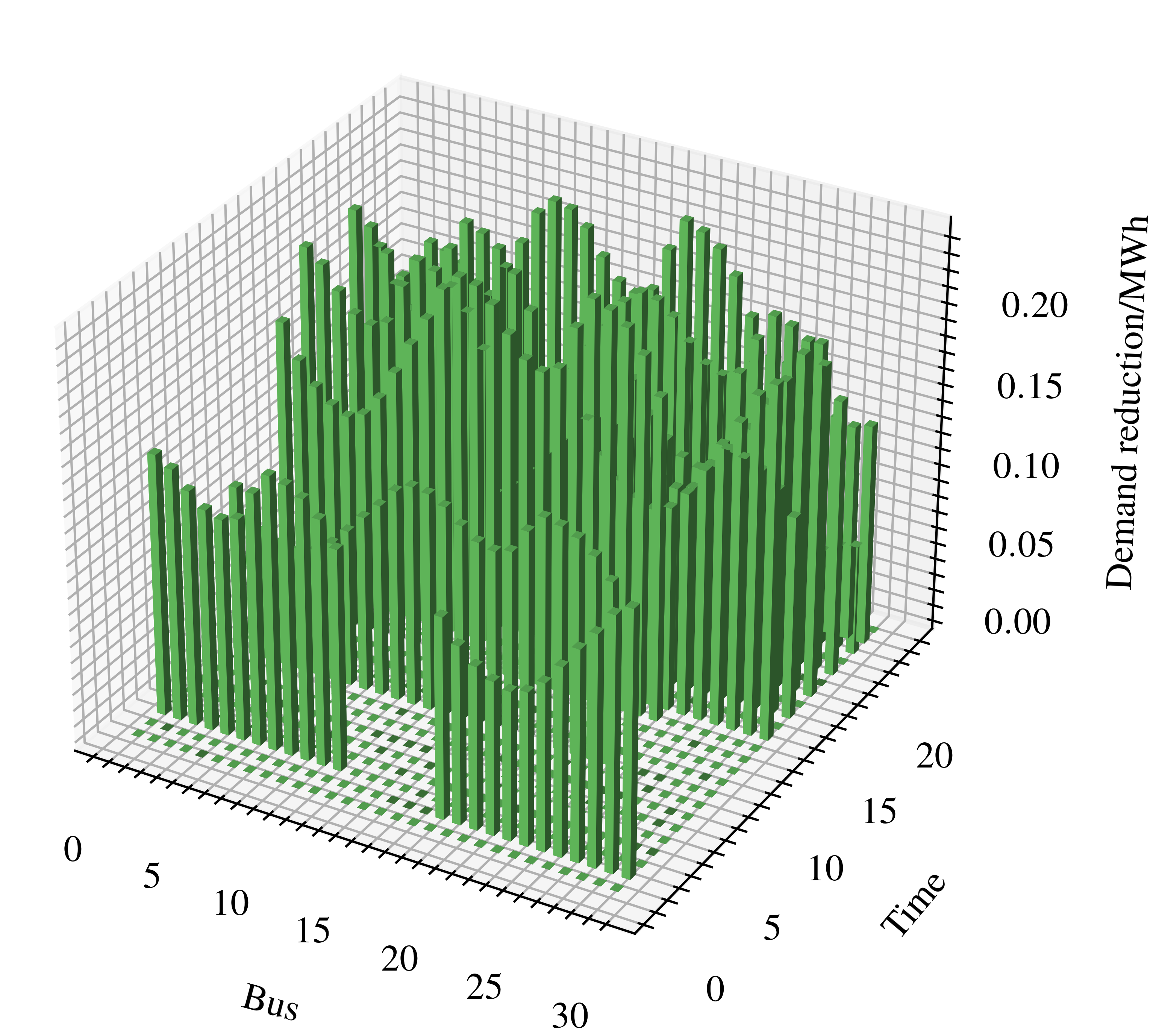}\label{fig: c30 dr a}}
  \subfigure[Demand response reduction analysis under the CA-EM in the 118-bus system.]{\includegraphics[scale=0.50]{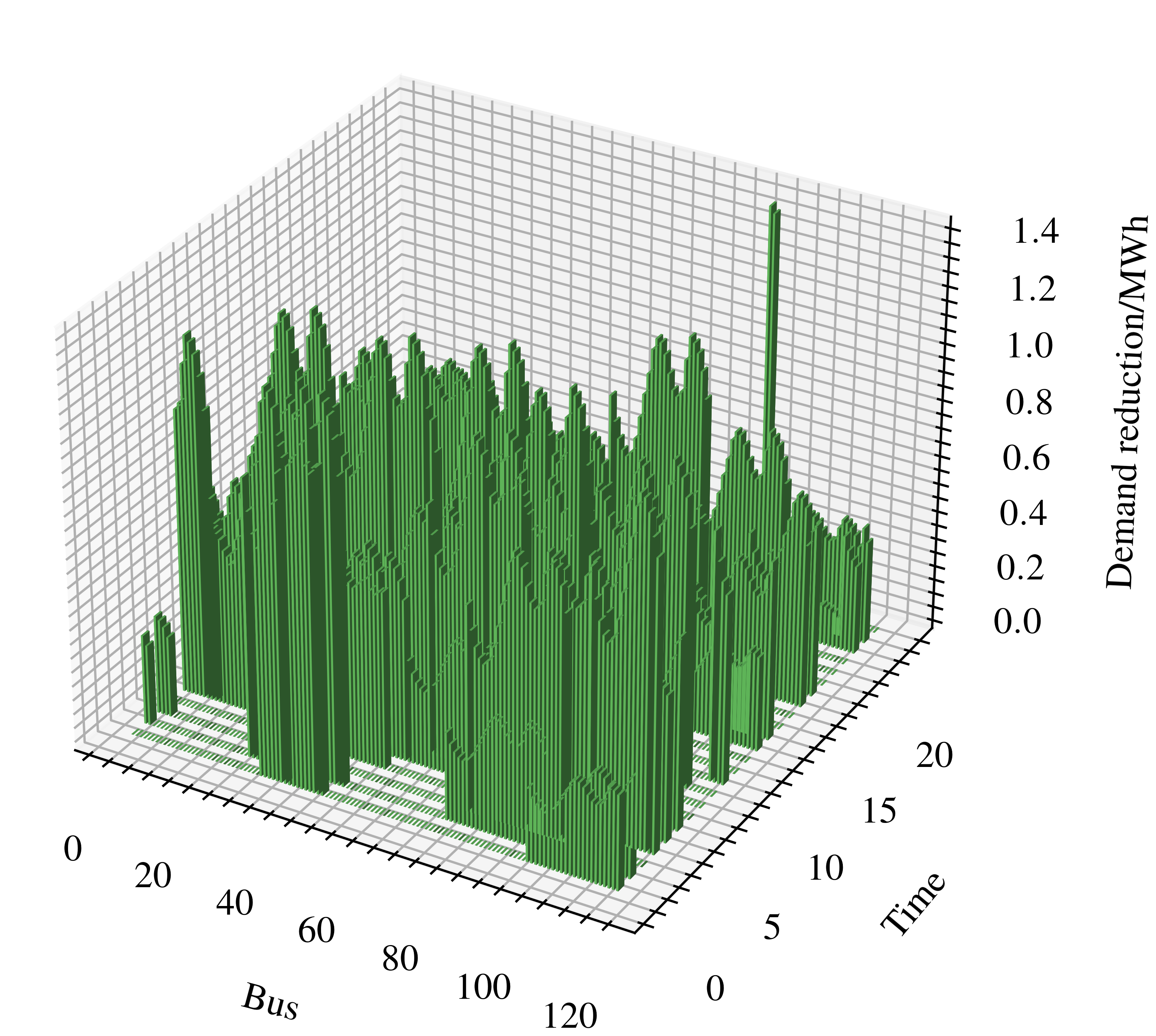}\label{fig: c30 dr b}}
  \caption{Demand response reduction analysis on different scheduling models in different systems.}
  \label{fig: c30 dr}
\end{figure}

\begin{figure}[ht]
  \centering
  \subfigure[Temporal carbon emission with ES charging and discharging for bus 2 in the 30-bus system.]{\includegraphics[scale=0.9]{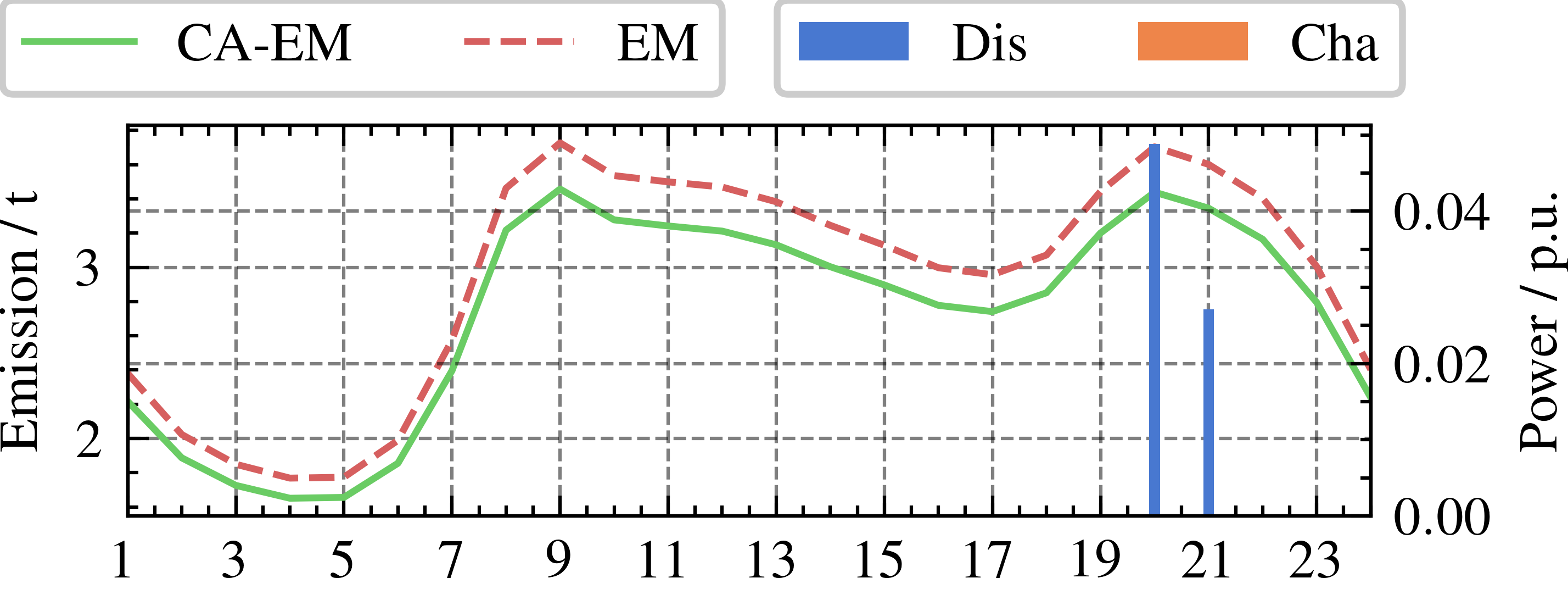}\label{fig: carbon emission a}}
  \subfigure[Temporal carbon emission with ES charging and discharging for bus 2 in the 118-bus system.]{\includegraphics[scale=0.9]{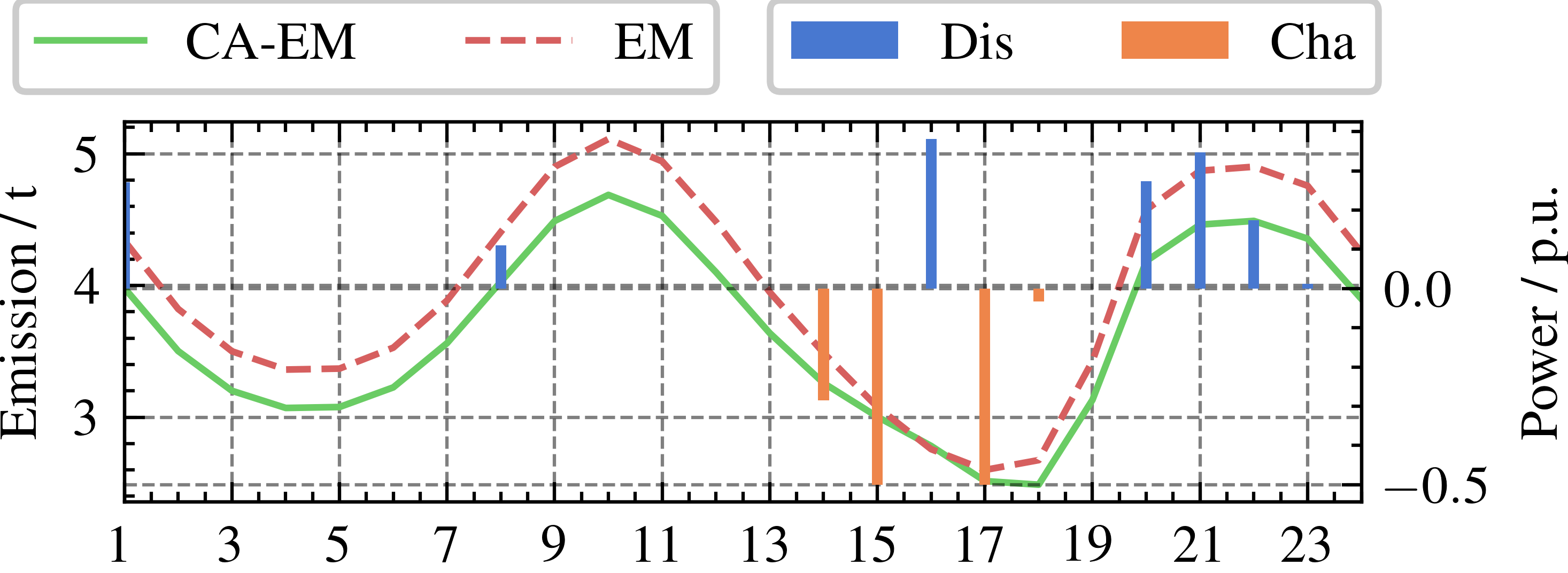}\label{fig: carbon emission b}}
  \caption{Temporal carbon emission with ES charging and discharging for bus 2 analysis.}
  \label{fig: carbon emission}
\end{figure}

Fig.~\ref{fig: c30 dr a} further presents each bus's demand side load reduction during each period in the 30-bus CA-EM model, {\rv verifying the effectiveness of incentivizing the carbon reduction potential by demand response}.
Fig.~\ref{fig: carbon emission a} compares the temporal carbon emissions with ES discharging and charging of bus 2 in CA-EM and EM with similar changing trends. When bus carbon emission is higher, the carbon emission reduction of CA-EM is more due to the blocked carbon price mechanism.

After visualizing the emission reduction effectiveness, Tab.~\ref{tab: c30 oper cost} compares the operation performance of EM and CA-EM numerically from the perspectives of social welfare, demand utility, carbon emission, and load consumption in the 30-bus system.
\begin{table}[ht]
  \renewcommand{\arraystretch}{1.3}
  \centering
  \caption{Operation comparison of different scheduling models in the 30-bus system.}
  \begin{tabular}{ccccc}
    \hline
    Models & Welfare(\$) & Utility(\$) & Emission(tCO$_2$) & Load(MWh) \\
    \hline
    EM & 31264.77 & 80859.13 & 1212.69 & 1386.43 \\ 
    CA-EM & 32553.29 & 79563.61 & 1128.45 & 1350.79 \\
    Change & +4.12\% & -1.60\% & -6.95\%  & -2.57\% \\
    \hline
  \end{tabular}
  \vspace{1ex}
  
  {\raggedright We note that the social welfare of EM is calculated in the way that social welfare without carbon awareness ignores the corresponding carbon cost. \par}
  \label{tab: c30 oper cost}
\end{table}

As illustrated in Tab.~\ref{tab: c30 oper cost}, CA-EM increases the total social welfare by incentivizing the carbon reduction potential on the demand side by reducing the load consumption by 35.64 MWh. It sacrifices the user load by reducing \$1295.52 utility but improves the system's low carbon operation by reducing 84.24 tCO$_2$ emissions. CA-EM leverages the reduction of 1.60\% load utility and 2.57\% load consumption to achieve 6.95\% carbon emission reduction, demonstrating its carbon reduction effectiveness in the 30-bus system.

\subsubsection{Case 118-bus Results Analysis}

\begin{figure}[ht]
  \centering
  \includegraphics[scale=0.9]{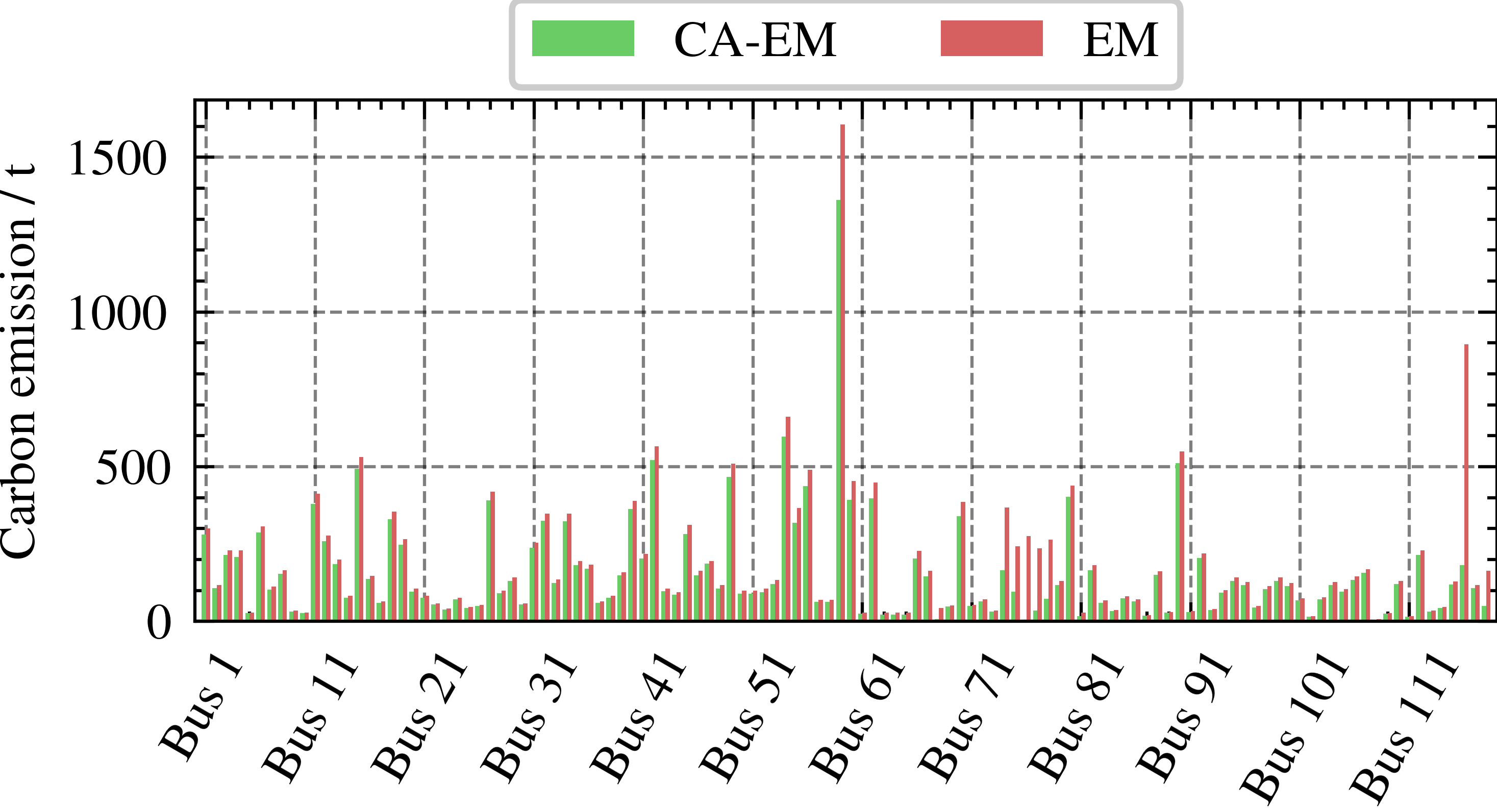}
  \caption{The node carbon emission analysis under different scheduling models in the 118-bus system.}
  \label{fig: c118 node bci}
\end{figure}

Similar to the analysis of the 30-bus system, for the 118-bus system, Fig.~\ref{fig: c118 node bci} compares the node carbon emission rates of CA-EM and EM. The proposed CA-EM can also reduce each bus's carbon emission effectively. Emission-intensive bus emission reductions are higher due to the proposed blocked carbon price mechanism on the demand side.

Fig.~\ref{fig: c30 dr b} further presents each bus's demand side load reduction during each period in the 118-bus CA-EM model, {\rv verifying the same effectiveness of demand response}. And Fig.~\ref{fig: carbon emission b} compares the temporal carbon emissions of bus 2 with ES in CA-EM and EM, verifying the carbon mitigation. Compared to Fig.~\ref{fig: carbon emission b}, ES exerts a more obvious effect by charging during low emissions and discharging during high emissions except time 16. Then Tab.~\ref{tab: c118 oper cost} also compares the operation performance of EM and CA-EM numerically from the perspectives of social welfare, demand utility, carbon emission, and load consumption in the 118-bus system.
\begin{table}[ht]
  \renewcommand{\arraystretch}{1.3}
  \centering
  \caption{Operation comparison of different scheduling models in the 118-bus system.}
  \begin{tabular}{ccccc}
    \hline
    Models & Welfare(\$) & Utility(\$) & Emission(tCO$_2$) & Load(MWh) \\
    \hline
    EM & 763139.78 & 1744722.14 & 22085.21 & 29598.33 \\ 
    CA-EM & 910215.07 & 1730705.60 & 18460.70 & 28948.63 \\
    Change & +19.27\% & -0.80\% & -16.41\%  & -2.19\% \\
    \hline
  \end{tabular}
  \label{tab: c118 oper cost}
\end{table}

As illustrated in Tab.~\ref{tab: c118 oper cost}, CA-EM achieves higher social welfare and lower emission with lower load utility and consumption. It leverages the reduction of 0.80\% load utility and 2.19\% load consumption to achieve 16.41\% carbon emission reduction, demonstrating its higher carbon reduction effectiveness in the 118-bus system.

\subsubsection{Benefits of Sparse Neural Networks}

SNN features fewer parameters through the sparse training of algorithm \ref{algo: cs-sgd}. So its equivalent MILP form will also exhibit fewer continuous and integer variables according to \eqref{eq: big-m}, improving the solving efficiency of the data-driven CA-EM. To verify the benefits of SNN, Tab.~\ref{tab: benefits of snn} compares the emission and solving time of CA-EM models with SNN and NN learning models.

\begin{table}[ht]
  \renewcommand{\arraystretch}{1.3}
  \centering
  \caption{Comparison of different neural networks in CA-EM scheduling.}
  \begin{tabular}{cccc}
    \hline
    Models & System & Emission (tCO$_2$) & Solving time(s)  \\
    \hline
    CA-EM with NN  & 30-bus & 1129.58 & 0.121s  \\ 
    CA-EM with SNN & 30-bus & 1128.45 & 0.058s \\
    CA-EM with NN  & 118-bus & 18461.85 & 28.943s \\ 
    CA-EM with SNN & 118-bus & 18460.70 & 1.387s \\
    \hline
  \end{tabular}
  \label{tab: benefits of snn}
\end{table}
As illustrated in Tab.~\ref{tab: benefits of snn}, CA-EM with SNN generates a similar carbon emission to the CA-EM with NN but achieves 2.08x solving time acceleration in the 30-bus system 20.86x acceleration in the 118-bus system, verifying the solving efficiency of SNN with energy-saving benefits. The above sparsity design is aligned with climate change mitigation in ML applications \cite{Kaack2022}. 

\subsection{Sensitivity Analysis} \label{sub: sa}

Then this part further conducts sensitivity analysis on the varying carbon prices impact of the proposed CA-EM model in the 30-bus and 118-bus systems. The following Fig.~\ref{fig: price sensitivity} compares the carbon emission and DR capacity with the increase of blocked carbon price. 

\begin{figure}[ht]
  \centering
  \subfigure[Carbon price sensitivity analysis in the 30-bus system.]{\includegraphics[scale=0.93]{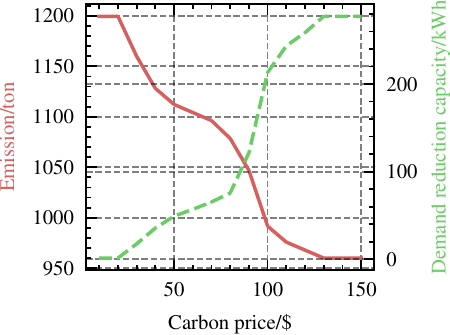}\label{fig: price sensitivity a}}
  \subfigure[Carbon price sensitivity analysis in the 118-bus system.]{\includegraphics[scale=0.93]{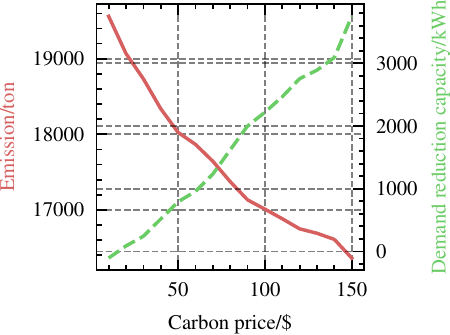}\label{fig: price sensitivity b}}
  \caption{Carbon price sensitivity analysis.}
  \label{fig: price sensitivity}
\end{figure}

For the 30-bus system of Fig.~\ref{fig: price sensitivity a}, when the carbon price is lower than \$20 /tCO$_2$ or higher than \$130 /tCO$_2$, the carbon price does not influence the carbon emission. During the interval from \$20 /tCO$_2$ to \$130 /tCO$_2$, the emission reduces fast at first, slows down, reduces fast again, and slows down finally; in contrast, the demand reduction capacity increases slowly at first, speeds up later and slows down later. This trend demonstrates that the initial load reduction capacity can generate a more obvious emission reduction performance with higher benefits. For the 118-bus system of Fig.~\ref{fig: price sensitivity b}, with the increase of carbon price, the demand reduction capacity increases steadily; in contrast, carbon emission reduces fast at first and slows down later, also demonstrating the higher benefits of initial load reduction. {\rv The above verifies the impact of carbon price setting on the proposed CA-EM model and provides the reference for the price setting.}

\section{Conclusion} \label{sec: conclusion}

This paper leverages ML and MILP to unlock the carbon reduction potential from the demand side. It proposes the compact constraint learning approach with SNN to learn the complex CEF models, transforms them into equivalent MILP constraints, and designs the blocked carbon price mechanism. Encoding the SNN-based constraints and price mechanism in the energy management model formulates the data-driven CA-EM model for emission mitigation. The case study verifies that 1) SNN features comparable approximation accuracy with dense NN but fewer parameters, improving the solution efficiency; 2) the CA-EM can incentivize the load reduction to mitigate the carbon emission effectively. {\rv The carbon-electricity coordinated optimization is limited to the transmission networks, which can be extended to the distribution networks and integrated energy systems for generalization, and future work can further consider incentivizing the carbon reduction potential of the emerging electrical vehicles and data centers.}

{\rv
\appendix

\subsection{Neural Network Tuning} \label{apx: nn tuning}

For the rationality of the hyperparameter setting, we leverage the grid search approach to tune the parameters of the neural networks, especially the number of neurons, as shown in Fig.~\ref{fig: param tuning}. With the increase of hidden layer neuron number, the NN mean score rises shapely at first and flattens later. The orange stars in both figures present the neuron number selection, where further the increase of neurons does not improve the score much but will increase the computational complexity with more variables and constraints in \eqref{eq: cef snn}. This demonstrates the rationality of the hyperparameter setting for SNN and NN in case studies.

\begin{figure}[ht]
  \centering
  \subfigure[Parameter tuning for case 30.]{\includegraphics[scale=0.93]{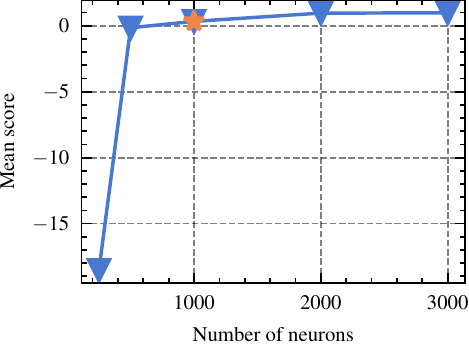}}
  \subfigure[Parameter tuning for case 118.]{\includegraphics[scale=0.93]{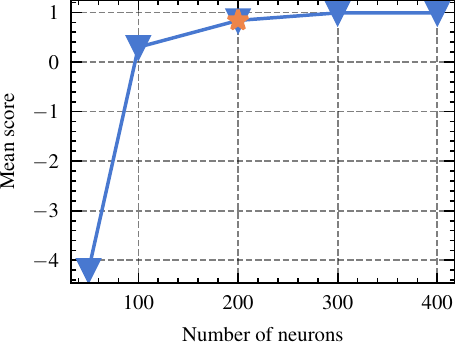}}
  \caption{{\rv Parameters tuning of neural networks in different cases.}}
  \label{fig: param tuning}
\end{figure}

}
\ifCLASSOPTIONcaptionsoff
  \newpage
\fi

\begin{IEEEbiography}[{\includegraphics[width=1in,height=1.25in,clip,keepaspectratio]{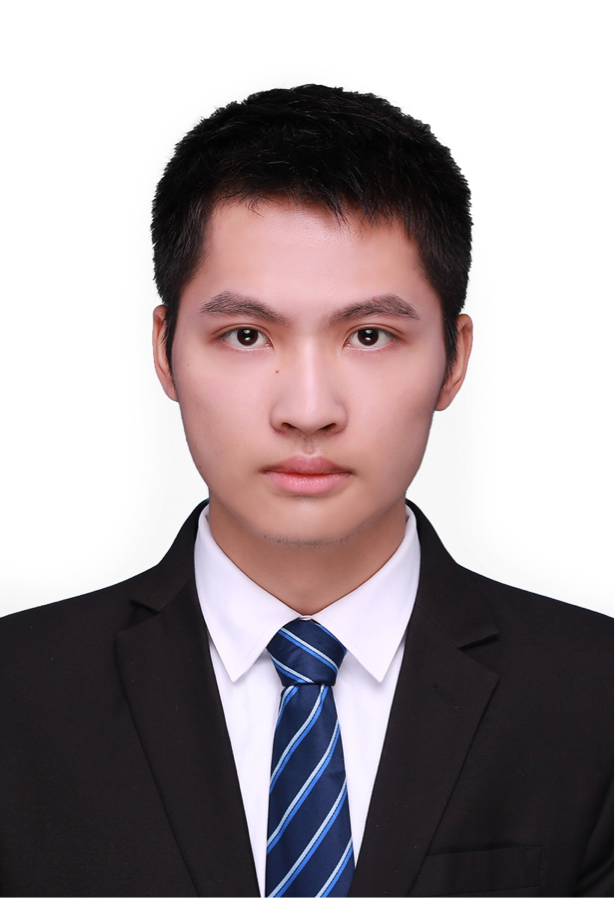}}]{Linwei Sang}(S'20) 
  received the M.S. degree from the School of Electric Engineering, Southeast University, China in 2021. 
  
  He is currently pursuing the Ph.D. degree in the Tsinghua-Berkeley Shenzhen Institute, Tsinghua University, Shenzhen, China. His research includes machine learning application in smart grid, the control of the distributed energy, and demand side resource management. 
\end{IEEEbiography}

\begin{IEEEbiography}[{\includegraphics[width=1in,height=1.25in,clip,keepaspectratio]{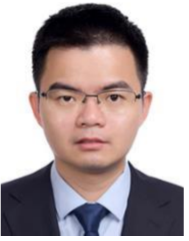}}]{Yinliang Xu}(SM'19)
  received the B.S. and M.S. degrees in control science and engineering from the Harbin Institute of Technology, Harbin, China, in 2007 and 2009, respectively, and the Ph.D. degree in electrical and computer engineering from New Mexico State University, Las Cruces, NM, USA, in 2013.
  
  He is currently an Associate Professor with Tsinghua-Berkeley Shenzhen Institute, Tsinghua Shenzhen International Graduate School, Tsinghua University, Beijing, China. His research interests include distributed control and optimization of power systems, renewable energy integration, and microgrid modeling and control. 
\end{IEEEbiography}

\begin{IEEEbiography}[{\includegraphics[width=1in,height=1.25in,clip,keepaspectratio]{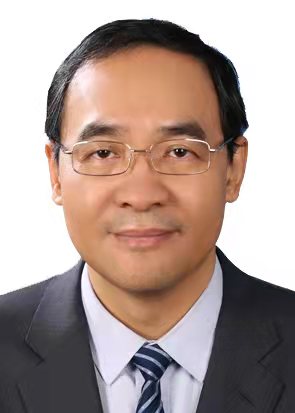}}]{Hongbin Sun}(Fellow, IEEE) received his double B.S. degrees from Tsinghua University in 1992, the Ph.D from Dept. of E.E., Tsinghua University in 1996. 
  
  He is now ChangJiang Scholar Chair professor and the director of energy management and control research center, Tsinghua University. He also serves as the editor of the IEEE TSG, associate editor of IET RPG, and member of the Editorial Board of four international journals and several Chinese journals. His technical areas include electric power system operation and control with specific interests on the Energy Management System, Automatic Voltage Control, and Energy System Integration.
\end{IEEEbiography}

\clearpage

\end{document}